\documentclass{jfm}

\usepackage{graphicx}
\usepackage{amsmath}
\usepackage{xcolor}
\usepackage{newtxtext}
\usepackage{newtxmath}
\usepackage{natbib}
\usepackage{threeparttable}
\usepackage{subfigure}
\usepackage{subcaption}
\usepackage{mleftright}
\usepackage{amssymb,bbding,oplotsymbl,MnSymbol}
\usepackage{relsize}
\usepackage{multirow}
\usepackage{float}
\usepackage{tikz} 
\usepackage{siunitx}
\usepackage[hyphens]{url}
\usepackage{hyperref}
\hypersetup{breaklinks=true}
\hypersetup{
  colorlinks = true,
  urlcolor  = blue,
  citecolor = blue,
}

\definecolor{maroon}{rgb}{0.64, 0.08, 0.18}	

\definecolor{d_cyan}{rgb}{0.0, 0.45,0.74}

\newcommand{\RomanNumeralCaps}[1]

\usepackage{ulem}
\title{Vortex ring formation from the interaction of a cavitation bubble with a confined air bubble: experiments and a timing criterion}

\author{Charul Gupta\aff{1},
Yashwant Singh\aff{1},
 Lakshmana D Chandrala\aff{1},
 Harish N Dixit\aff{1,2},
 \and Badarinath Karri\aff{1}
  \corresp{\email{badarinath@mae.iith.ac.in}},
}

\affiliation{\aff{1}Department of Mechanical \& Aerospace Engineering, Indian Institute of Technology Hyderabad, India
\aff{2}Centre for Interdisciplinary Programs, Indian Institute of Technology Hyderabad, India}

\begin{document}
\maketitle

\begin{abstract}
We study vortex ring formation arising from the interaction between a cavitation bubble and a confined air bubble in a cylindrical blind hole, using high-speed shadowgraphy imaging. As the cavitation bubble grows above the hole, it drives a downward flow that compresses the air bubble at the base. The air bubble subsequently expands, expelling the overlying liquid column upward as a coherent slug; impact of this slug on the far boundary of the collapsing cavitation bubble produces a vortex ring. Parametric experiments across the dimensionless stand-off distance $\mathcal{H} = h/R_{\max}$ and the air bubble fill fraction $\mathcal{B} = (d_\text{hole} - d_\text{top})/d_\text{hole}$ identify three regimes: (i)~liquid column impact during collapse, producing a vortex ring ($\mathcal{H} \lesssim 0.5$, $\mathcal{B} \lesssim 0.5$); (ii)~late impact near the end of collapse (large $\mathcal{H}$); and (iii)~direct air bubble impact after bypassing the liquid column (large $\mathcal{B}$), with neither (ii) nor (iii) producing a ring. Two one-dimensional models, based on the Rayleigh--Plesset equation and isentropic air bubble expansion, predict the liquid column impact location and its speed $U_\text{lc}$, respectively. A dimensionless timing parameter $\Pi = (h + R_{\max}) / (U_\text{lc} \cdot t_\text{cav}/2)$, comparing the liquid column travel time to the cavitation collapse half-period, distinguishes the three regimes: ring formation occurs for $1 \lesssim \Pi \lesssim 1.5$. The ring propagates from the hole at an initial speed of $5$~m/s, decelerating quadratically, and breaks apart via azimuthal instabilities at $Re \approx 4500$.
\end{abstract}

\begin{keywords}
Authors should not enter keywords on the manuscript, as these must be chosen by the author during the online submission process and will then be added during the typesetting process (see \href{https://www.cambridge.org/core/journals/journal-of-fluid-mechanics/information/list-of-keywords}{Keyword PDF} for the full list). Other classifications will be added at the same time.
\end{keywords}

{\bf MSC Codes } {\it(Optional)} Please enter your MSC Codes here


\section{\label{sec:Intro}Introduction}
The collapse of a cavitation bubble near a boundary is one of the most energetic and well-studied events in fluid mechanics. In the canonical configuration, a single bubble collapsing near a rigid planar wall, the collapse drives a high-speed liquid jet directed toward the wall, which is responsible for cavitation erosion and the basis of many practical applications. A central question in the field is how the nature and geometry of the boundary modifies this jet: redirecting it, suppressing it, or replacing it entirely with a qualitatively different coherent structure. The present study addresses a configuration in which the boundary contains a confined gas pocket, and shows that this seemingly minor modification produces a fundamentally different outcome: rather than a jet, the collapse event generates a coherent vortex ring that propagates away from the surface. Understanding the mechanism behind this transition, from jet formation to vortex ring formation, is the central aim of the paper.

Cavitation is a fundamental physical phenomenon characterised by the formation, growth, and violent collapse of vapor bubbles in a liquid when the local pressure drops below the vapor pressure. It arises in a wide range of engineering and biological contexts, including hydraulic machinery, marine propellers \citep{arndt1981cavitation}, and biomedical and microfluidic devices \citep{sarraf2022fundamentals}. Although cavitation can cause significant physical damage near solid surfaces, the same dynamics can be exploited constructively in applications such as targeted drug delivery \citep{Coussios2008}, ultrasonic cleaning \citep{Yusof2016} and wastewater treatment \citep{mancuso2020critical}. A particularly relevant geometry is a blind hole, i.e., a cylindrical cavity closed at one end, which arises in ultrasonic cleaning of engineering components, in cavitation near porous or rough surfaces, and in bubble-driven transport in confined microfluidic channels. The interaction between a cavitation bubble and a gas pocket trapped in such a cavity has received little attention despite its practical importance.

The theoretical foundation for cavitation bubble dynamics was established by \cite{rayleigh1917}, who derived the equation governing spherically symmetric bubble oscillation in an infinite incompressible liquid. The dynamics near solid surfaces were clarified by the seminal experimental study of \cite{benjamin1966collapse} and the theoretical model of \cite{plesset1971}, both of which showed that collapse near a rigid wall drives a high-speed jet directed toward the wall. Subsequent studies \citep{lauterborn1975,tomita1984collapse,vogel1989optical,zhang1993,philipp1998cavitation,lindau2003cinematographic} established that the jet speed and direction depend critically on the stand-off distance $\mathcal{H}$, defined as the ratio of the bubble-centre-to-surface distance to the maximum bubble radius $R_{\max}$. In particular, \cite{vogel1989optical} and \cite{lindau2003cinematographic} showed that both a jet and a counter-jet can form over certain ranges of $\mathcal{H}$.
\cite{gonzalez2011cavitation} showed that collapse in a narrow gap produces either jetting or neutral collapse depending on the stand-off distance. \cite{mnich2024single} demonstrated that a cavitation bubble generated in the stagnation region of a wall jet migrates upstream and collapses away from the wall. Across all these configurations, the stand-off distance emerges as the primary geometrical control parameter.

The nature of the bounding surface also plays a decisive role. Near a free surface \citep{chahine1977interaction, blake1981}, collapse deforms the interface into a liquid spike while a micro-jet forms inside the bubble directed away from the free surface. \cite{gisbon1980growth} reported similar jet-away-from-surface behaviour near flexible boundaries. A comprehensive review of bubble behaviour near various boundary types is given by \cite{blake1987cavitation}. Near elastic surfaces, \cite{brujan2001dynamics} documented bubble splitting, jet formation in either direction, and penetration into the elastic material depending on surface compliance and stand-off distance.

While most earlier studies focused on bubble collapse near planar surfaces, later investigations explored more complex geometries such as through-holes and confined cavities. \cite{Khoo2005} and \cite{Lew2007} demonstrated that collapse near a through-hole in a submerged solid produces a high-speed jet passing across the hole; \cite{Karri2011} examined the dependence of this jet on viscosity and stand-off distance. \cite{Karri2012,Karri2012b} subsequently examined a configuration in which one end of the through-hole was exposed to air and the other was immersed in water, and observed the formation of a pair of jets. \cite{GonzalezAvila2015} extended this to tapered holes, generating high-speed micro-jets and sprays over a wide range of Weber numbers. \cite{heidary2024robust} showed that jet formation near a capillary tube produces a jet whose diameter is governed by the dimensionless capillary diameter and the stand-off distance. Notably, \cite{Lew2007} observed vortex ring formation during single-bubble collapse near a through-hole, but attributed it to the circulation induced by the jet passing across the hole, which is a mechanism distinct from what we report here. All of these studies considered the collapse of a single bubble near a fixed geometric feature.

More recently, attention has turned to the interaction between cavitation bubbles and neighbouring gas bubbles. \cite{Bai2011} reported counter-jet formation from such an interaction. \cite{Pain2012} showed that when a gas bubble is encapsulated in a silicone oil droplet, the interaction drives jet formation inside the gas bubble, with dynamics depending strongly on inter-bubble separation. \cite{Goh2014} placed a hemispherical air bubble between the cavitation bubble and a solid wall and found that the jet direction is governed by the separation between the two bubbles. Interactions with solid particles \citep{poulain2015particle,teran2018interaction} show that particles continue to move toward the bubble both during and after collapse. In all of these cases, the gas bubble is free to expand in any direction. The qualitatively new feature of the present configuration is that the gas bubble is confined in a blind hole, so it can only expand in one direction, i.e., upward into the liquid column above it. This enforced directionality is what makes a coherent liquid slug possible, and ultimately what leads to vortex ring formation.

The present study demonstrates that a cavitation bubble generated just above a blind hole containing a confined air bubble produces, under certain conditions, a coherent vortex ring that propagates away from the surface. This represents a new mechanism for vortex ring generation: the growing cavitation bubble compresses the confined air, which re-expands and ejects the overlying liquid as a directed slug; when this slug impacts the far boundary of the cavitation bubble during its collapse phase, an axisymmetric impulse generates azimuthal vorticity that rolls up and detaches as a ring. The outcome is governed by two dimensionless parameters, the stand-off distance $\mathcal{H} = h/R_{\max}$ and the air bubble fill fraction $\mathcal{B} = (d_\text{hole} - d_\text{top})/d_\text{hole}$, and by a timing criterion $\Pi$ that compares the liquid slug travel time to the cavitation collapse half-period. To our knowledge, the only prior study of this geometry is that of \cite{Kauer2018}, who focused on air bubble removal rather than flow structure generation. The hole aspect ratio (depth to diameter) is fixed at 3. Three intermediate processes, namely air bubble compression and expansion, penetrating bubble formation, and liquid column motion, are characterised across all experiments. Two one-dimensional models, based on the Rayleigh–Plesset equation, are developed: the liquid column model predicts the impact point of the liquid column, while the expansion model of the air bubble predicts the liquid column speed and the timing criterion.

The paper is organised as follows. The experimental setup and methodology are described in \S\ref{sec:setup}. \S\ref{sec:key_cases} presents the parametric study and three representative cases. The theoretical model for the cavitation bubble and air bubble dynamics is developed in \S\ref{sec:theory}. Conclusions are provided in \S\ref{sec:discussion}.

\section{\label{sec:setup}Experimental methodology}
\begin{figure}
 \begin{minipage}{0.69\textwidth}
 \centering
   \subfigure[]{\label{fig:schematic1}
   \includegraphics[trim = 0mm 0mm 0mm 0mm, clip, angle=0,width=\textwidth]{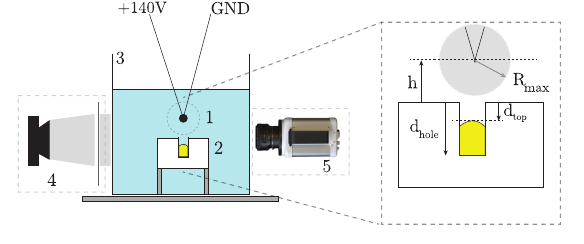}
   }\\
   \subfigure[]{\label{fig:schematic2}
   \includegraphics[trim = 0mm 0mm 0mm 0mm, clip, angle=0,width=0.85\textwidth]{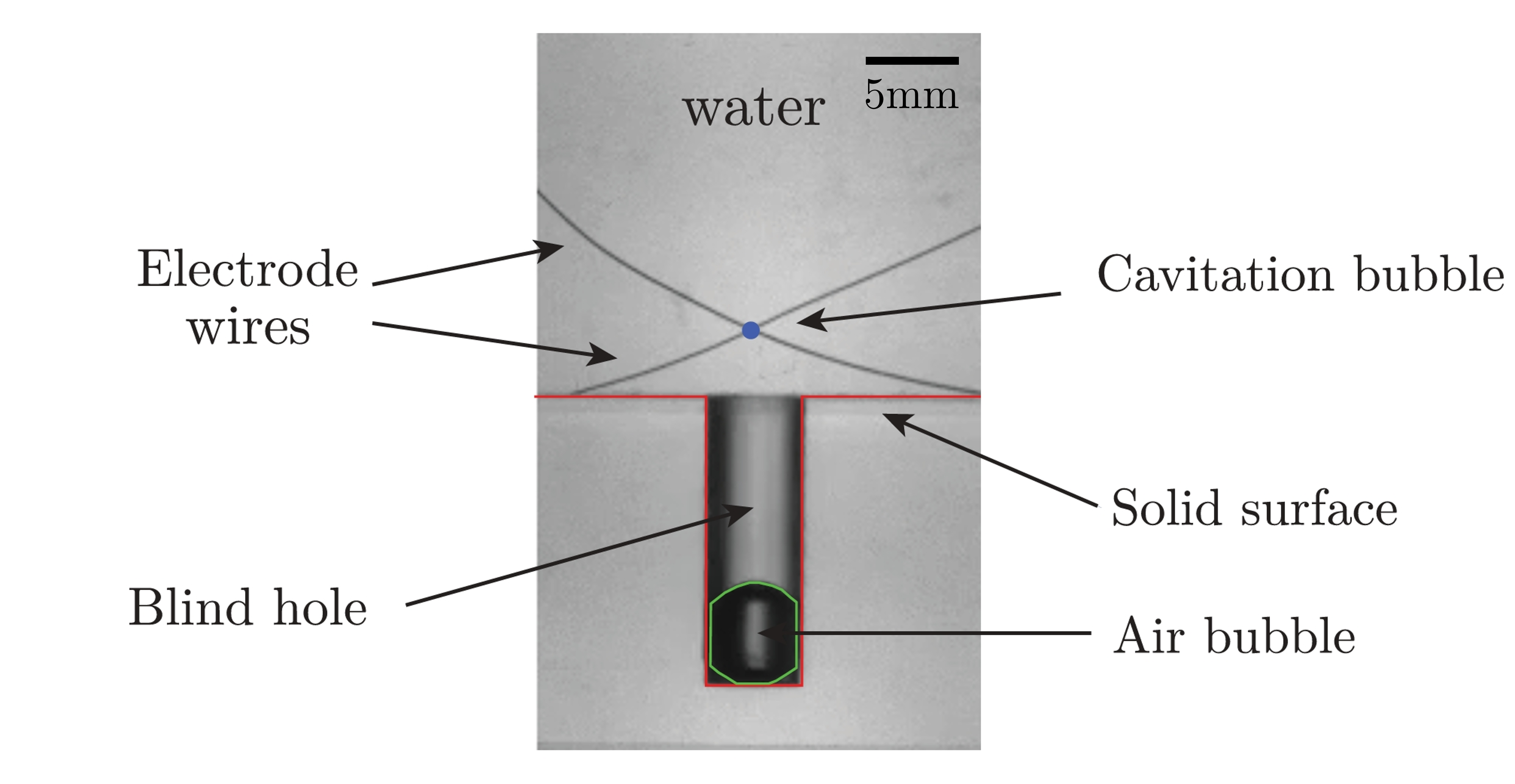}
   }
 \end{minipage}
 \begin{minipage}{0.3\textwidth}
   \centering
   \subfigure[]{\label{fig:Ring-Track}
   \includegraphics[trim = 0mm 10mm 140mm 0mm, clip, angle=0,width=\textwidth]{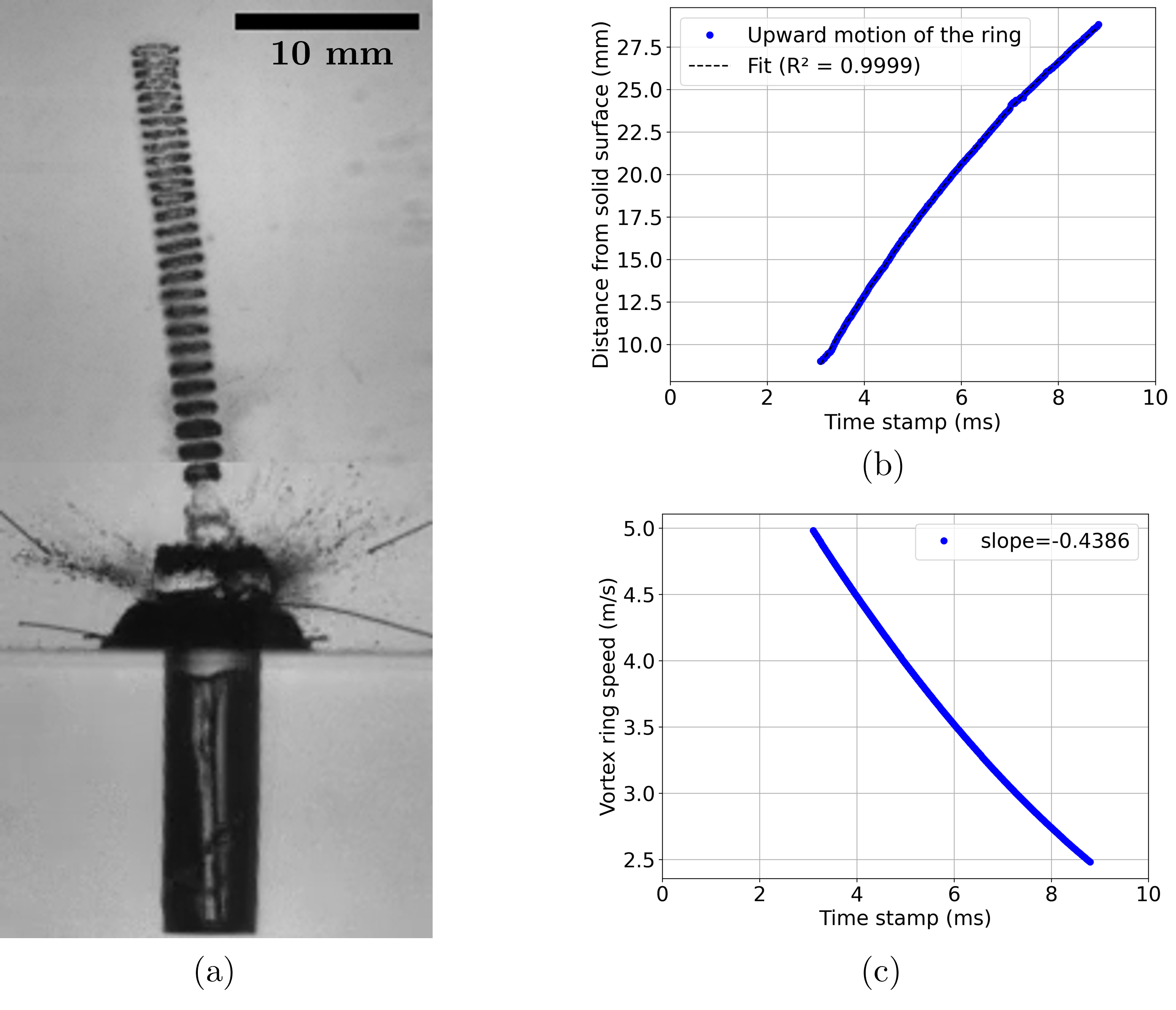}
   }
 \end{minipage}
 \caption{(a)~Schematic of the experimental setup; numbered components are described in the text. The inset defines the key geometric parameters $h$, $R_\text{max}$, $d_\text{hole}$, and $d_\text{top}$. (b)~Representative experimental image (spatial resolution 5.8--13.4~pixels/mm). Red lines: top surface of the acrylic block; green contour: air bubble inside the blind hole; blue-filled circle: electrode contact point. (c)~Composite image of the vortex ring trajectory for Case~1 ($\mathcal{H}=0.44$, $\mathcal{B}=0.37$), formed by superimposing frames at 500~$\mu$s intervals from $t = 3225$~$\mu$s to $t = 10000$~$\mu$s.}
\label{fig:exp_schematic}
\end{figure} 
Figure~\ref{fig:schematic1} shows the experimental setup, which comprises: (1) contacting copper wires with an electronic circuit, (2) a solid acrylic block with a blind hole at its top centre, (3) an acrylic tank, (4) an illumination system, and (5) a high-speed imaging system. The tank ($250~\text{mm} \times 250~\text{mm} \times 250~\text{mm}$) is filled to the brim with distilled water. The solid acrylic block ($100~\text{mm} \times 100~\text{mm} \times 50~\text{mm}$) is placed at the centre of the tank and fixed to prevent movement during experiments. The block is positioned sufficiently far from the free surface to eliminate free-surface effects. The solid block has a cylindrical blind hole of a diameter of 5 mm and a depth of $d_{\text{hole}} = 15$ mm centrally located at its top surface. An air bubble (highlighted in yellow in Figure~\ref{fig:schematic1}) is introduced at the base of the hole with a controlled height $d_\text{hole} - d_\text{top}$ using a syringe attached to a nylon tube; care is taken to ensure axisymmetry with respect to the hole axis.
\begin{table}
\centering
\caption{Summary of the three representative cases. $R_\text{max}$: maximum cavitation bubble radius; $t_\text{cav}$: total bubble lifetime from generation to complete collapse.}
\begin{tabular}{c c c c c}
\hline
\textbf{Key Cases} & \textbf{$\mathcal{B}$} & \textbf{$\mathcal{H}$} & \textbf{Maximum Radius (mm)} &
\textbf{$t_\text{cav}$ ($\mu$s)}\\ \hline
Case 1 & 0.37 & 0.44 & 7.3 & 3300 \\ \hline
Case 2 & 0.37 & 0.72 & 6.2 & 2200 \\ \hline
Case 3 & 0.63 & 0.24 & 8.2 & 3750 \\ 
\end{tabular}
\label{tab1:Summary}
\end{table}

A low-voltage spark-discharge method comprising an electronic circuit and two contacting electrodes \citep{goh2013low} is then used to generate a cavitation bubble on top of the blind hole. The circuit comprises three distinct sub-circuits to perform key functions: (i) charging circuit, (ii) discharging circuit, and (iii) spark circuit. The charging circuit comprises capacitors that allow a constant voltage to be applied and maintained across the electrodes. The discharging circuit performs two key functions: (1) allowing excess voltage to be discharged, which helps maintain a constant voltage across the electrodes, and (2) discharging the residual voltage remaining after the spark. The spark circuit uses a MOSFET to apply the constant voltage across the electrodes, which eventually leads to spark generation. These electrodes are connected with two tinned copper wires arranged in a point contact just above the blind hole at a known distance ($h$) from the solid surface. The location of the contact point is accurately controlled using an XYZ traverse mechanism (not shown in Figure~\ref{fig:schematic1}). Figure~\ref{fig:schematic2} shows the real-time view of the region of interest, where a pair of electrodes is located at the top of the blind hole and the air bubble is contained at its bottom end.

The circuit is controlled via a LabVIEW GUI through a National Instruments DAQ system. To generate a bubble, the capacitors are first charged to 140~V; the MOSFET is then triggered to apply this voltage across the electrodes, producing a spark and nucleating the cavitation bubble; finally, the discharging circuit removes residual voltage. 
The phenomenon is recorded using a Photron Nova S9 high-speed camera with a 100~mm macro lens at 20\,000--40\,000~frames/s ($160 \times 400$ to $512 \times 760$~pixels; exposure time: 1--4~$\mu$s, spatial resolution 5.8--13.4~pixels/mm). Uniform back-lit illumination is provided by a white LED source with a diffuser sheet. A delay generator triggers the camera at spark ignition, ensuring consistent capture from bubble generation through complete collapse.

Two dimensionless parameters characterise the interaction. The air bubble fill fraction $\mathcal{B} = (d_{\mathrm{hole}}-d_{\mathrm{top}})/d_{\mathrm{hole}}$ measures the fraction of the hole occupied by the air bubble (the hole is cylindrical, so height and volume fractions are equivalent). The stand-off distance $\mathcal{H} = h/R_{\mathrm{max}}$ is the ratio of the electrode height above the block surface to the maximum bubble radius; $R_{\max}$ is measured from the experimental images at the instant of maximum bubble growth by equating the projected bubble area to that of a circle. Experiments were conducted over the range $\mathcal{H} < 1$ and $\mathcal{B} < 1$. Figure~\ref{fig:Ring-Track} shows a composite image of the vortex ring trajectory at $\mathcal{H}=0.44$ and $\mathcal{B}=0.37$ within the specified range of $\mathcal{H}-\mathcal{B}$ space, constructed by superimposing frames at 500~$\mu$s intervals between 3225~$\mu$s and 10000~$\mu$s.

The maximum bubble radius was held within $R_{\max} = 6$--$8.5$~mm by fixing the discharge voltage at 140~V; multiple repeats at each parameter setting confirmed reproducibility. Figure~\ref{fig:All_data} consolidates all experimental data in the $\mathcal{H}$--$\mathcal{B}$ space. Three representative cases, highlighted with large filled markers in Figure~\ref{fig:All_data} and listed in Table~\ref{tab1:Summary}, are selected to span the three distinct regimes and are discussed in detail in \S\ref{sec:key_cases}.

\subsection{Data analysis}
High-speed image sequences were pre-processed in ImageJ to enhance contrast. Because multiple processes (air bubble, penetrating bubble, liquid column, cavitation bubble) occur simultaneously in the same field of view, direct segmentation of the full image is not straightforward. Images were therefore cropped at the block surface to separate processes inside and outside the blind hole, and each region was segmented independently. Segmentation used the WEKA trainable segmentation plugin for ImageJ, which implements a random forest classifier trained on user-provided pixel masks; the classifier employs an ensemble of feature kernels (Gaussian smoothing, edge detection, local mean and median) and provides accurate boundary extraction. The segmented images from the two regions were recombined and analysed using custom MATLAB scripts to extract bubble and interface positions as functions of time. 

\section{\label{sec:key_cases}Parametric study}
\begin{figure}
\centering
\includegraphics[trim = 0mm 0mm 0mm 0mm, clip, angle=0,width=0.9\textwidth]{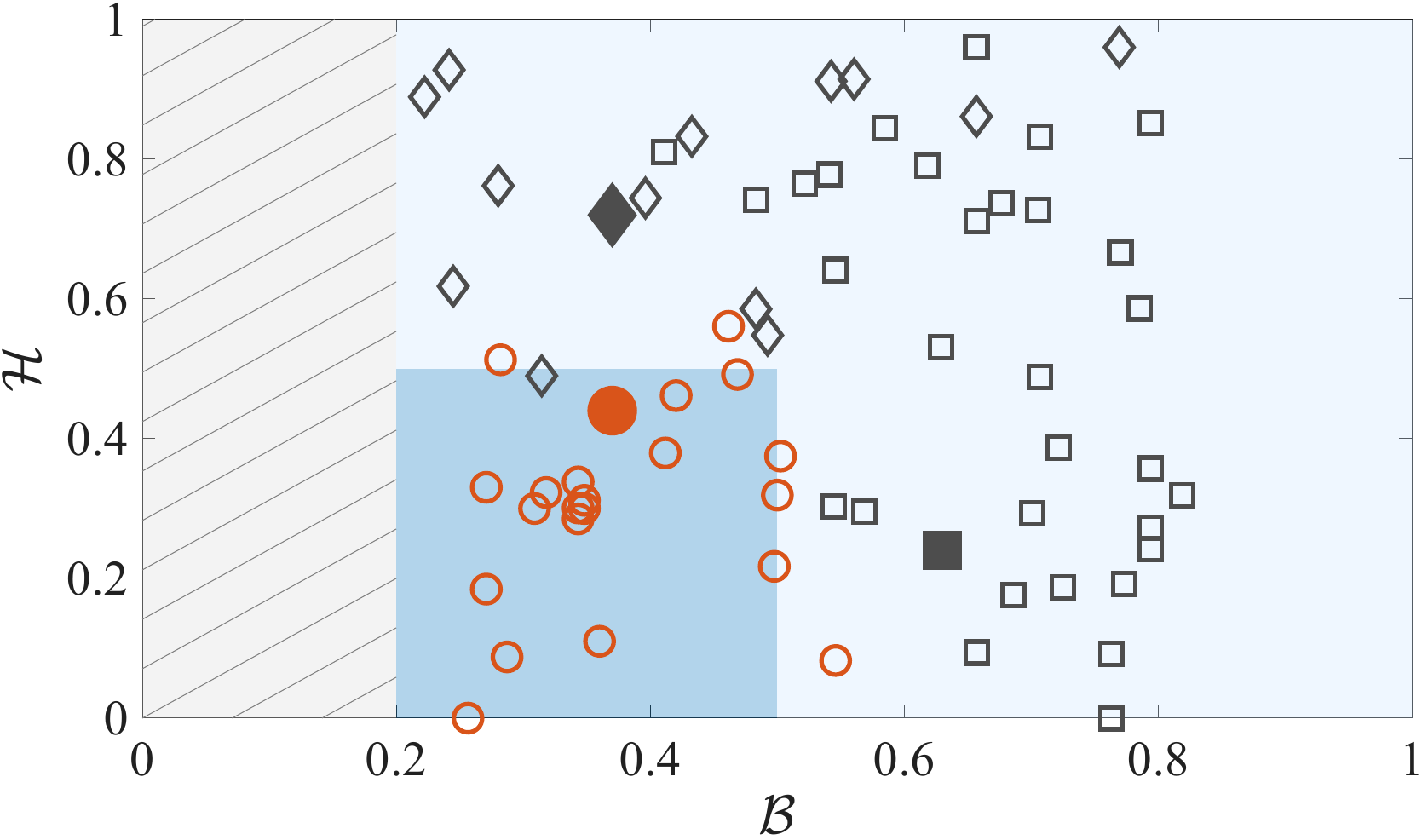}
\caption{Regime map of all experimental data in the $\mathcal{H}$--$\mathcal{B}$ parameter space. Orange circles~(\scalebox{2}{\textcolor{orange}{$\circ$}}): liquid column impact during the collapse phase, producing a vortex ring (L-B Impact); dark grey diamonds~($\color{darkgray}{\large\boldsymbol{\Diamond}}$): late liquid column impact near the end of collapse, no ring (Late Impact); dark grey squares~($\color{darkgray}{\large\boldsymbol{\Box}}$): direct air bubble impact on the far cavitation boundary, no ring (B-B Impact). The grey-hatched region ($\mathcal{B}<0.2$) indicates parameter combinations where the minimum air bubble size could not be achieved experimentally. The three representative cases discussed in \S\ref{sec:key_cases} are highlighted with larger filled markers.}
\label{fig:All_data}
\end{figure}

Figure~\ref{fig:All_data} consolidates all experimental data in the $\mathcal{H}$--$\mathcal{B}$ parameter space. Three distinct interaction categories are identified: (i) liquid column impact on the far cavitation boundary during collapse, producing a vortex ring (L-B Impact, orange circles); (ii) late liquid column impact near the end of collapse, producing no ring (Late Impact, grey diamonds); and (iii) direct air bubble impact on the far boundary, producing no ring (B-B Impact, grey squares). These three outcomes are illustrated through three representative cases: Case~1 ($\mathcal{H} = 0.44$, $\mathcal{B}=0.37$), Case~2 ($\mathcal{H} = 0.72$, $\mathcal{B}=0.37$), and Case~3 ($\mathcal{H} = 0.24$, $\mathcal{B}=0.63$), described in the following subsections. 
\subsection{Case 1 - Impact between Liquid column and Cavitation bubble boundary (L-B Impact)}\label{sec:case-1}
Case~1 ($\mathcal{H} = 0.44$, $\mathcal{B} = 0.37$, $R_{\max} = 7.3$~mm) is the canonical vortex-ring-forming case. The electrode contact point is located $h = 3.2$~mm above the block surface. 

\begin{figure}
\centering
\includegraphics[trim = 0mm 0mm 0mm 0mm, clip, angle=0,width=0.8\textwidth]{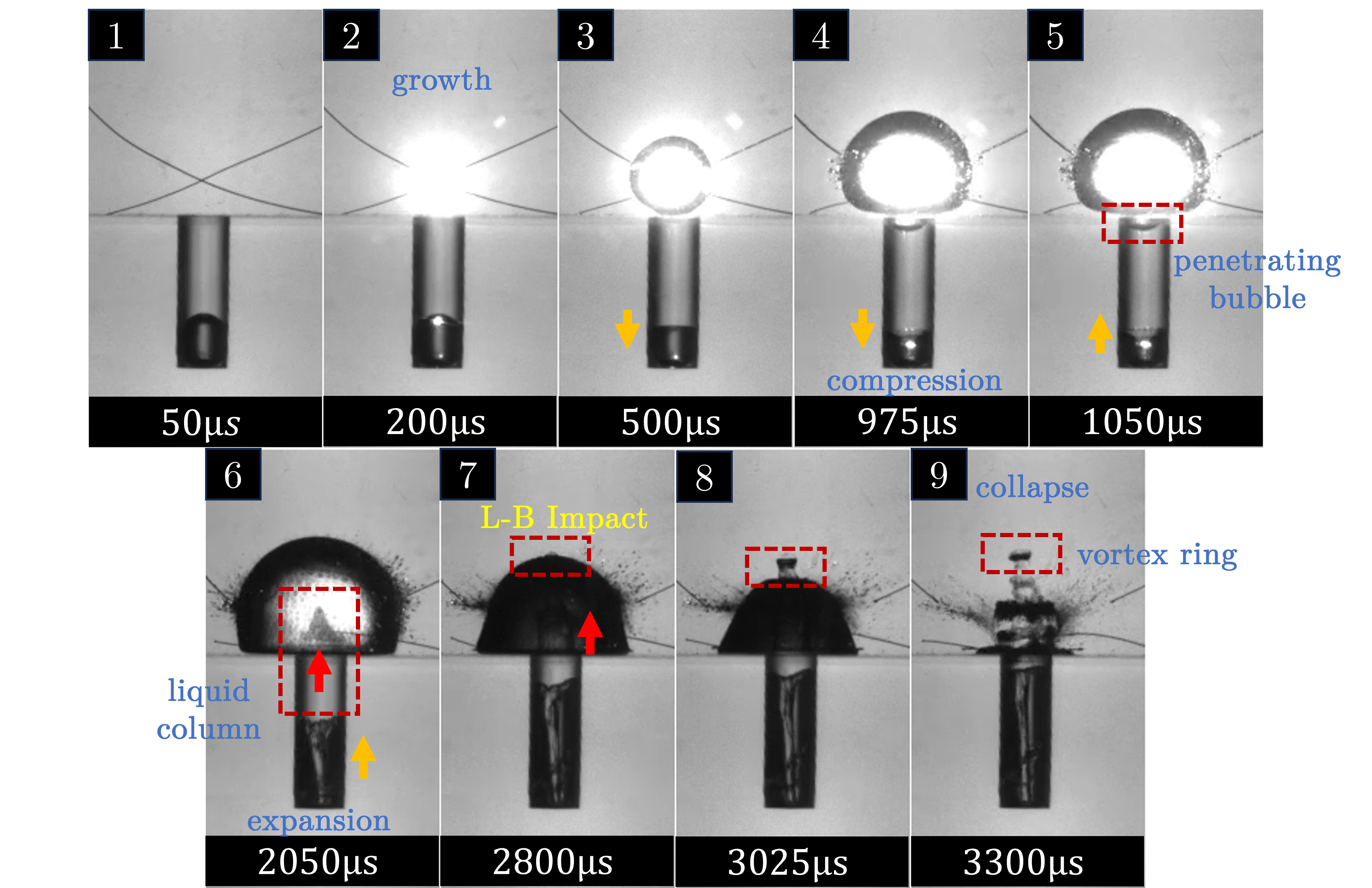}
\caption{Image sequence for Case~1 ($\mathcal{H} = 0.44$, $\mathcal{B} = 0.37$, $t_\text{cav} = 3300~\mu$s). Image~1: bubble nucleation. Image~2: onset of air bubble compression (downward yellow arrow). Image~3: entry of the penetrating bubble into the blind hole. Image~4: maximum penetrating bubble depth. Image~5: onset of air bubble expansion (upward yellow arrow) and exit of the penetrating bubble (maroon box). Image~6: maximum cavitation bubble radius; the conical tip of the liquid column is visible (maroon box). Image~7: liquid column impact on the far cavitation boundary (L-B impact; upward red arrow shows column motion). Image~8: vortex ring pinching off from the collapsing bubble (maroon box). Image~9: complete collapse. The complete temporal sequence of the process for this case is shown in \href{https://people.iith.ac.in/hdixit/Cavitation_supplementary.html}{supplementary} movie 1.}
\label{fig:case-1_phenomena}
\end{figure} 
Figure \ref{fig:case-1_phenomena} displays a temporal sequence of all events during the lifetime of the cavitation bubble ($3300~\mu$s, shown in Image 9). The first image (Image 1) shows the generation of the cavitation bubble, and the last image (Image 9) depicts its complete collapse. The following three intermediate processes occur between these two events: (a) air bubble compression (Image 2) and expansion (Image 5), (b) entry (Image~3) and exit (Image~5) of the penetrating bubble, and (c) impact of the liquid column with the far boundary of the cavitation bubble (Image 7). The penetrating bubble is the portion of the cavitation bubble that enters and exits the blind hole at its top end (Images~3 and~5). The liquid column denotes the column of the liquid between the cavitation bubble and the air bubble (see Image 6). Image~8 shows the vortex ring pinching off from the collapsing bubble boundary, which is the key outcome of the present study. Notably, the standard downward jet toward the solid surface, which is the expected outcome of cavitation bubble collapse near a rigid wall, is absent here: the confinement of the air bubble redirects the dynamics entirely, replacing the jet with a coherent ring propagating away from the surface.

\begin{figure}
\centering
\includegraphics[trim = 0mm 0mm 0mm 0mm, clip, angle=0,width=0.9\textwidth]{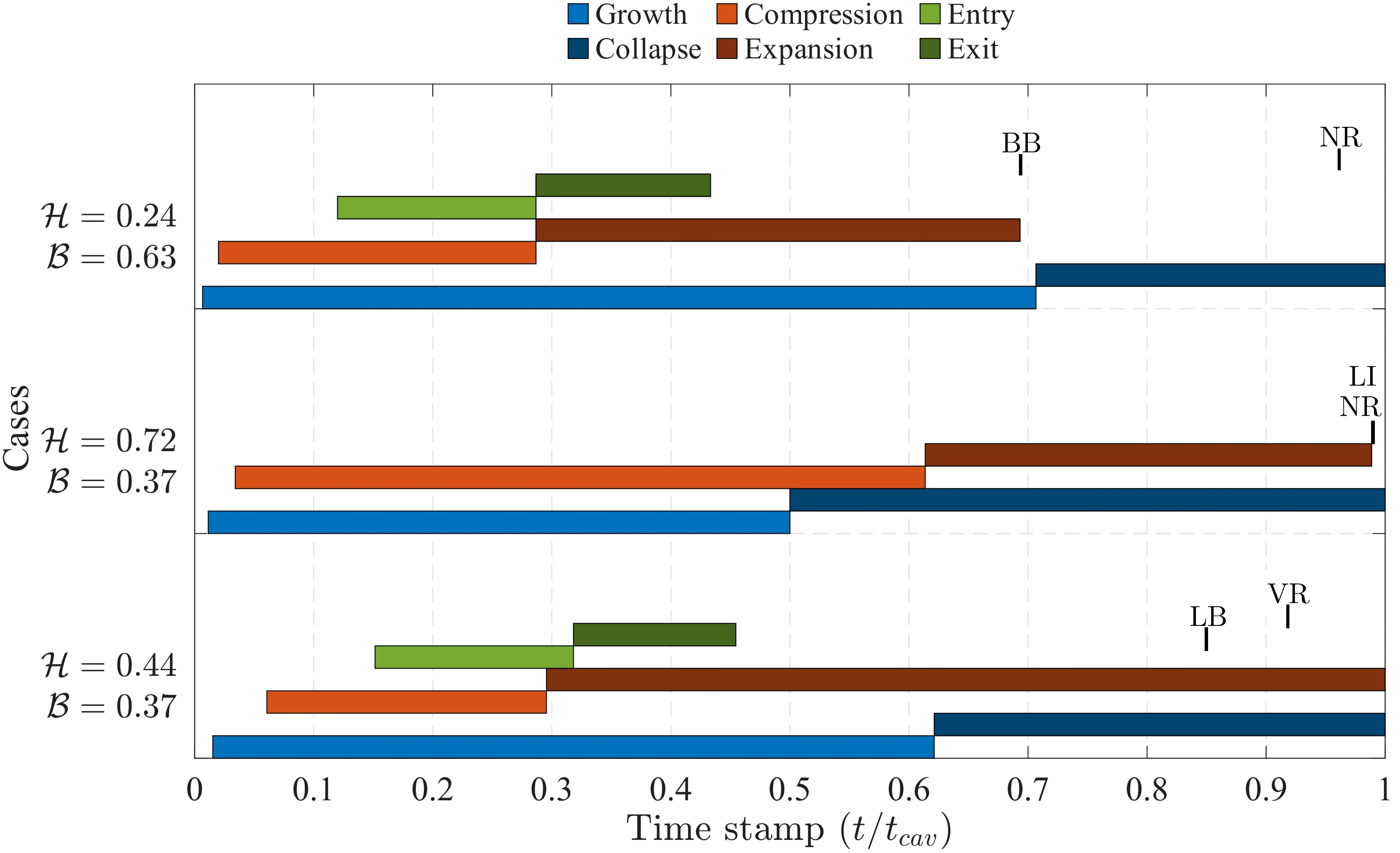}
\caption{Cavitation time map for the three representative cases. The $x$-axis is dimensionless time $t/t_\text{cav}$, where $t_\text{cav}$ is the total bubble lifetime for each case. Coloured bars indicate the duration of each process: blue, cavitation bubble growth; dark blue, collapse; orange, air bubble compression; dark orange, expansion; green, penetrating bubble entry; dark green, exit. Instantaneous events are shown as text annotations: `LB', liquid column-bubble boundary impact; `VR', vortex ring formation; `LI', late impact; `BB', air bubble impact on far boundary; `NR', no ring formed.}
\label{fig:Time_Stamp}
\end{figure}

The formation of the vortex ring depends on the timing of the impact of the liquid column with the collapsing bubble boundary; therefore, knowledge of the timings of these events is crucial. A cavitation time map is shown in Figure~\ref{fig:Time_Stamp}, which consolidates all events along with their corresponding timings to indicate their occurrence. For Case 1, $\mathcal{H} =0.44$ and $\mathcal{B}=0.37$, the key processes with their dimensionless time stamps are presented in the bottom panel of Figure~\ref{fig:Time_Stamp}. The lower pair of bars (blue and dark blue bars) represent the growth and collapse stages of the cavitation bubble (associated with Images 1, 6 and 9 in Figure~\ref{fig:case-1_phenomena}). Note that the duration of the growth stage is longer than that of the collapse stage; therefore, the bubble oscillation is asymmetric. This behaviour of cavitation bubbles has been reported in several studies, primarily in the vicinity of solid surfaces \citep{brennen2014cavitation}. The next pair of bars (orange and dark orange bars) denote the compression and expansion of the air bubble (Images 2 and 5), both of which begin during the growth stage of the cavitation bubble. The expansion of the air bubble persists until the complete collapse of the cavitation bubble (see the dark orange bar). The top two bars of this case in the bottom panel (green and dark green bars) denote the occurrence of the entry and exit of the penetrating bubble (see Images 3 and 5) within the blind hole. A detailed discussion of the air bubble dynamics is provided in \S\ref{sec:air_bubble}; the role of the penetrating bubble is described in \S\ref{sec:discussion}. The annotations `LB' and `VR' indicate that the liquid column impacts the far boundary of the cavitation bubble, consequently leading to the formation of a vortex ring. The annotation `LB', shown approximately midway through the collapse stage indicates the timing of the liquid column impact and the vortex ring is formed soon after (annotation `VR'). The vortex ring formed in this case is discussed in detail in \S\ref{sec:ring_trajectory}.
\subsection{Case 2 - Late Impact between Liquid column and Cavitation bubble boundary}\label{sec:case-2}
Case~2 ($\mathcal{H} = 0.72$, $\mathcal{B} = 0.37$, $R_{\max} = 6.2$~mm) has the same air bubble fill fraction as Case~1 but a significantly larger stand-off distance. The electrode is $h = 4.46$~mm above the surface. 

\begin{figure}
\centering
\includegraphics[trim = 0mm 0mm 0mm 0mm, clip, angle=0,width=0.75\textwidth]{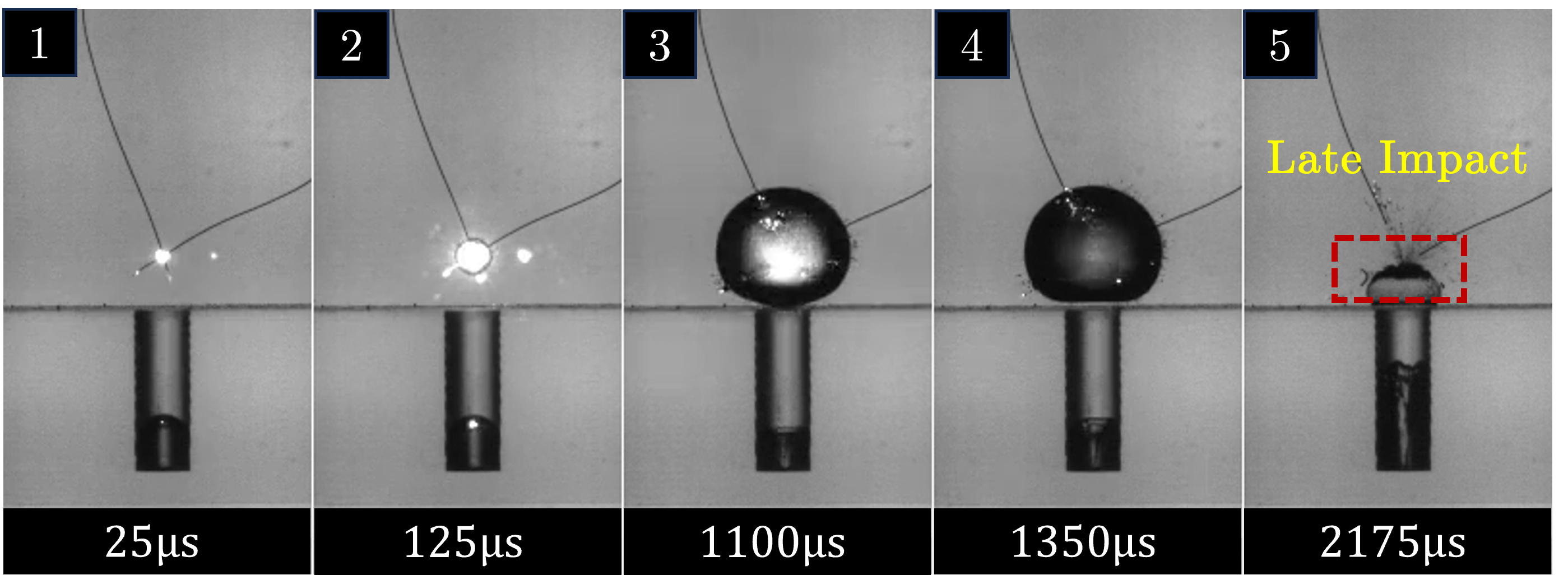}
\caption{Image sequence for Case~2 ($\mathcal{H} = 0.72$, $\mathcal{B} = 0.37$, $t_\text{cav} = 2200~\mu$s). Image~1: bubble nucleation. Image~2: onset of air bubble compression. Image~3: onset of air bubble expansion. Image~4: maximum cavitation bubble radius. Image~5: complete collapse; the red dashed box highlights the toroidal deformation of the liquid column caused by the collapsing jet. No vortex ring forms. See \href{https://people.iith.ac.in/hdixit/Cavitation_supplementary.html}{supplementary} movie 2 for the temporal evolution of the interaction in this case.}
\label{fig:case-2_phenomena}
\end{figure}

Figure~\ref{fig:case-2_phenomena} presents five images illustrating the processes that occurred between the growth and collapse stages of the cavitation bubble. The dimensionless time stamps for all processes are presented in the cavitation time map in the middle panel of Figure \ref{fig:Time_Stamp}. Figure~\ref{fig:case-2_phenomena} shows the five-image sequence spanning $t_\text{cav} = 2200~\mu$s. Two processes are observed: air bubble compression (Images~2--3) and expansion (Images~4--5), followed by outward motion of the liquid column (Image~5). The expanding air bubble drives the liquid column outward between 1350 and 2175~$\mu$s; however, the cavitation bubble collapses before the column reaches its far boundary (annotation `LI'). The annotation `NR' in the middle panel of Figure~\ref{fig:Time_Stamp} confirms that no ring forms.   

The cavitation time map (Figure~\ref{fig:Time_Stamp}, middle row) shows that growth and collapse are nearly symmetric in duration for Case~2, in contrast to the asymmetry seen in Cases~1 and~3, which are generated closer to the surface. Critically, the air bubble expansion begins only during the collapse stage. This delayed expansion means the liquid column arrives at the far cavitation boundary too late: the collapse is already dominated by the inward jet, which deforms the emerging column into a toroidal shape (red dashed box, Image~5 of Figure~\ref{fig:case-2_phenomena}) rather than pinching off a ring. No penetrating bubble forms ($\mathcal{H} = 0.72 > 0.5$), consistent with the linear relationship between the air-bubble compression height and $\mathcal{H}$, which also governs the penetration bubble height noted in \S\ref{sec:Pi}. 

\subsection{Case 3 - Impact between Air bubble and Cavitation bubble boundary (B-B Impact) }\label{sec:case-3}
\begin{figure}
\centering
\includegraphics[trim = 0mm 0mm 0mm 0mm, clip, angle=0,width=0.65\textwidth]{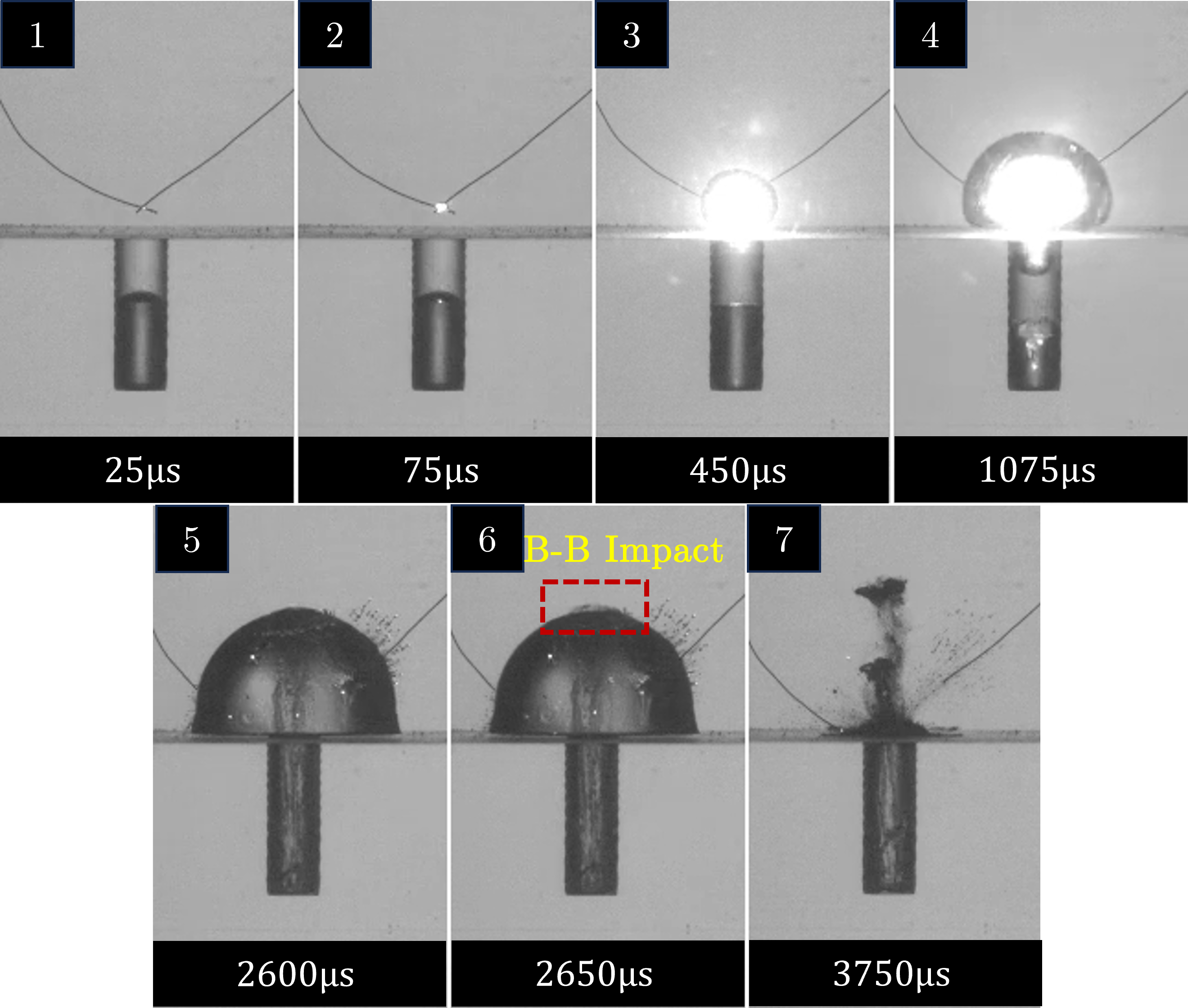}
\caption{Image sequence for Case~3 ($\mathcal{H} = 0.24$, $\mathcal{B} = 0.63$, $t_\text{cav} = 3750~\mu$s). Image~1: bubble nucleation. Image~2: onset of air bubble compression. Image~3: entry of the penetrating bubble. Image~4: onset of air bubble expansion and exit of the penetrating bubble. Image~5: maximum cavitation bubble radius. Image~6: direct impact of the expanding air bubble on the far cavitation boundary (B-B impact; red dashed box). Image~7: complete collapse; no vortex ring forms. The complete temporal sequence of the process for this case is shown in \href{https://people.iith.ac.in/hdixit/Cavitation_supplementary.html}{supplementary} movie 3.}
\label{fig:case-3_phenomena}
\end{figure} 
Case~3 ($\mathcal{H} = 0.24$, $\mathcal{B} = 0.63$, $R_{\max} = 8.2$~mm) is generated very close to the surface ($h = 1.96$~mm) with a large air bubble occupying 63\% of the hole.

Figure \ref{fig:case-3_phenomena} presents all processes using a sequence of seven images with their time stamps. The cavitation time map is also shown at the top panel in Figure \ref{fig:Time_Stamp}, with six bars representing the processes corresponding to the images in Figure \ref{fig:case-3_phenomena}. Images 1 and 7 depict the onset of cavitation bubble growth and the end of the collapse stage, respectively. The corresponding bars in the cavitation time map are coloured blue and dark blue. The two key phenomena, (i) air-bubble compression and expansion, and (ii) the entry and exit of the penetrating bubble, occurred in a manner similar to Case~1. In Figure \ref{fig:Time_Stamp} (top panel), the orange and dark orange bars indicate the compression of the air bubble and the subsequent expansion, respectively. These processes are shown in Images 2 and 4 in Figure \ref{fig:case-3_phenomena}. Image 5 represents the end of the growth stage, where the cavitation bubble reaches its maximum radius. The onset of the entry and exit of the penetrating bubble is shown in Images 3 and 4, with their corresponding green and dark green bars on the cavitation time map. In contrast to Case~1 where the liquid column strikes the far side of the cavitation bubble, here, the expanding air bubble impacts the cavitation bubble boundary.  
The annotation `BB' and Image 6 indicate the impact of the expanding air bubble on the far boundary of the cavitation bubble. This entire event did not result in the formation of the vortex ring, as indicated by the annotation `NR' on the cavitation time map and Image 7 in Figure \ref{fig:case-3_phenomena}. 

Unlike Case~2, both compression and expansion of the air bubble occur entirely during the growth stage (Figure~\ref{fig:Time_Stamp}, top row). However, the large $\mathcal{B} = 0.63$ means the liquid column above the air bubble is short, and the strongly compressed air bubble expands rapidly enough to overtake and fragment it. The air bubble itself strikes the far cavitation boundary (B-B impact, Image~6), causing only a weak interfacial roll-up without ring formation (Image~7). 
 
Having described the three distinct experimental outcomes, we now develop two one-dimensional theoretical models to characterise the dynamics of both the cavitation bubble and the confined air bubble, and to predict the impact location and speed of the liquid column that governs vortex ring formation.

\section{One-dimensional theoretical models}\label{sec:theory}
Two distinct models are developed to describe the interaction of the confined air bubble with the cavitation bubble. Both models require different inputs from the experiment. The first model, termed the `liquid-column model' and described in \S\ref{sec:liquid__column_model}, requires an accurate description of the cavitation bubble growth stage using the Rayleigh--Plesset framework. The Rayleigh--Plesset equation requires knowledge of the initial bubble radius, $R_\text{o}$, and initial pressure, $P_\text{o}$, which are used as initial conditions for the growth stage. The growth of the cavitation bubble pushes liquid into the blind hole, thus compressing the air bubble. Since the relationship between air bubble volume and pressure is not independently constrained, there is no \emph{a priori} reason to expect an accurate match of the compressed height of the air bubble. This is the limitation of the liquid-column model. To overcome this difficulty, we develop an alternate model which circumvents the growth phase of the cavitation bubble. This model, termed the `air bubble expansion model' and described in \S\ref{sec:air_bubble}, uses the minimum height of the air bubble at its most pressurised state, referred to as $z_{\text{in}}$, as a tuning parameter to match the expansion trajectory of the air bubble with the experiments. These two models give two different pieces of information about the interaction. The former provides information about when the liquid column impacts the cavitation bubble, while the latter provides a clear timing criterion based on the liquid slug speed.

\subsection{Liquid column model}\label{sec:liquid__column_model}
\subsubsection{Cavitation bubble dynamics}\label{sec:cav_model}
%
\begin{figure}
\centering
\subfigure[]{\label{fig:rp_case1}
\includegraphics[trim = 0mm 0mm 0mm 0mm, clip, angle=0,width=0.45\textwidth]{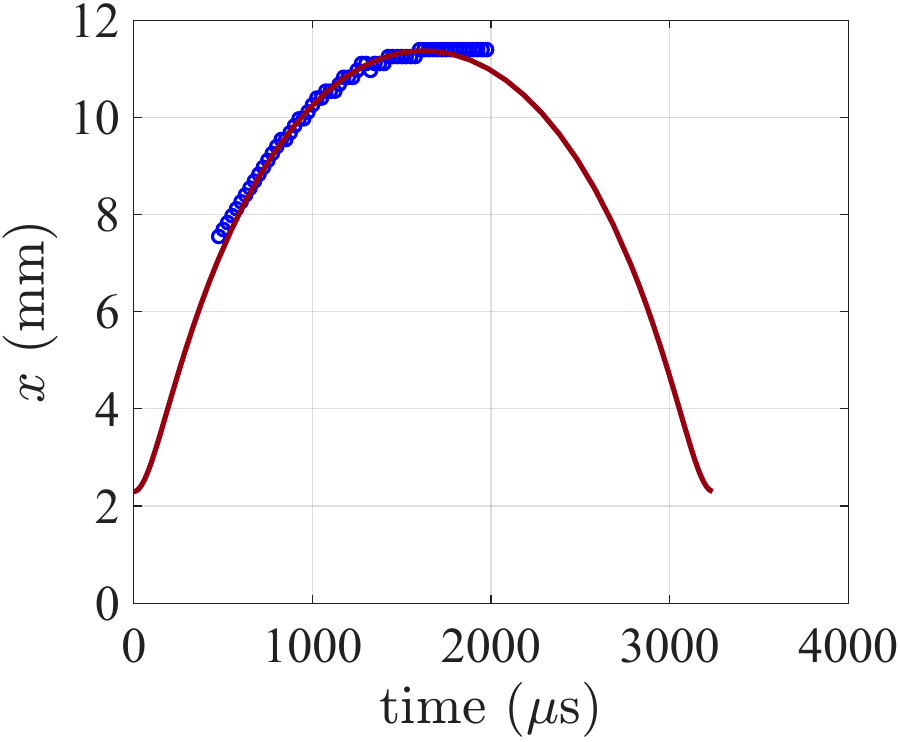}
}
\hspace{5mm}
\subfigure[]{\label{fig:rp_case2}
\includegraphics[trim = 0mm 0mm 0mm 0mm, clip, angle=0,width=0.45\textwidth]{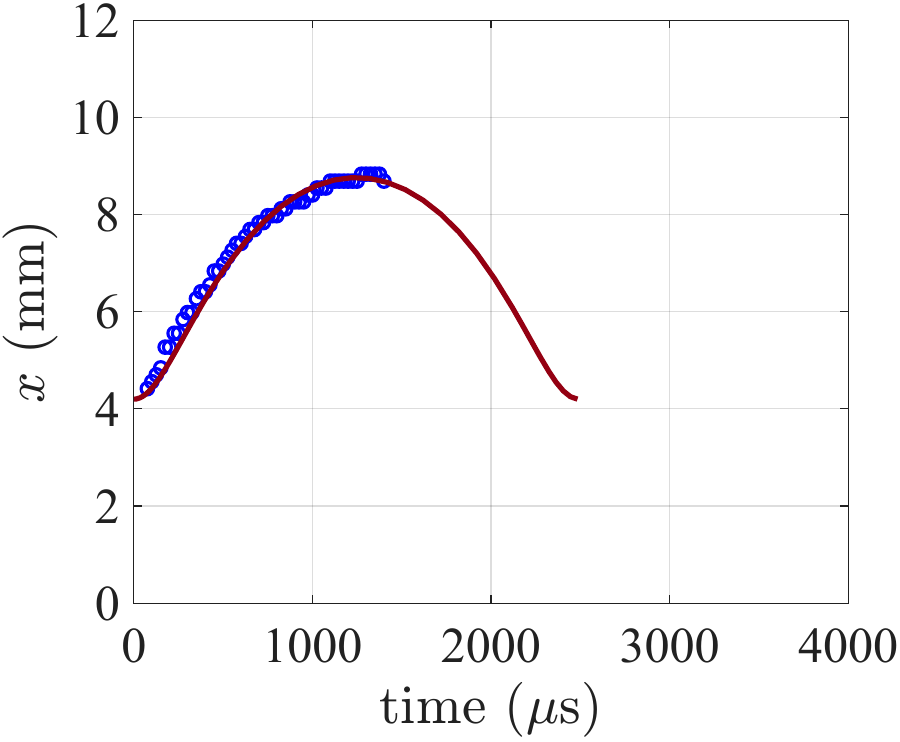}
}
\subfigure[]{\label{fig:rp_case3}
\includegraphics[trim = 0mm 0mm 0mm 0mm, clip, angle=0,width=0.45\textwidth]{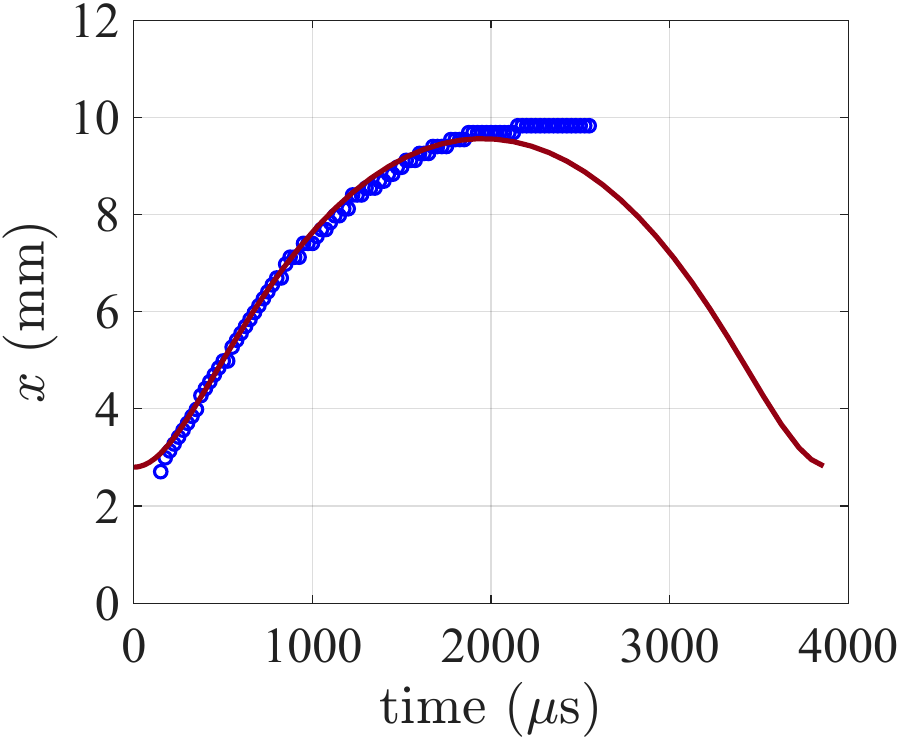}
}
\caption{Growth trajectory of the cavitation bubble: experiment (blue circles) versus the Rayleigh--Plesset model (maroon curve). Here $x$ is the distance of the far bubble boundary from the solid surface and $t = 0$ is the instant of bubble generation. The model is fitted by adjusting $P_\text{o}$ and $R_\text{o}$ to match the experimental growth stage. Fitted parameters: (a)~Case~1 ($\mathcal{H} = 0.44$, $\mathcal{B} = 0.37$): $P_\text{o} = 0.95\times10^6$~Pa, $R_\text{o} = 5.8$~mm; (b)~Case~2 ($\mathcal{H} = 0.72$, $\mathcal{B} = 0.37$): $P_\text{o} = 0.38\times10^6$~Pa, $R_\text{o} = 5.8$~mm; (c)~Case~3 ($\mathcal{H} = 0.24$, $\mathcal{B} = 0.63$): $P_\text{o} = 0.35\times10^6$~Pa, $R_\text{o} = 9.3$~mm.}
\label{fig:rp_comparison}
\end{figure}
The growth of the cavitation bubble is described using the Rayleigh--Plesset model \citep{rayleigh1917,plesset1949dynamics,poritsky1952collapse}. Assuming an incompressible surrounding medium with uniform properties (density $\rho_\text{l}$, kinematic viscosity $\nu_\text{l}$), the equation of motion for the bubble radius $R(t)$ reads:
\begin{equation}
 R\frac{d^2R}{dt^2} + \frac{3}{2} \left(\frac{dR}{dt}\right)^2 + \frac{4\nu_\text{l}}{R}\left(\frac{dR}{dt}\right) = \frac{p_\text{B}(t) - p_{\infty}}{\rho_\text{l}}, \label{eq:Rayleigh_Plesset}
\end{equation}
where $p_{\infty}$ is the ambient (atmospheric) pressure and $p_\text{B}$ is the internal pressure of the bubble. Assuming an isentropic process during growth (radial):
\begin{equation}
p_\text{B} = P_\text{o}\left(\frac{R_\text{o}}{R}\right)^{3k}, \label{eq:Isentropic-compression}
\end{equation}
where $P_\text{o}$ and $R_\text{o}$ are the bubble pressure and radius at the instant of generation, and $k = 1.4$ is the polytropic index. The parameters $P_\text{o}$ and $R_\text{o}$ are determined by iteratively fitting equation~\ref{eq:Rayleigh_Plesset} to the experimental growth trajectory for each case. Figure~\ref{fig:rp_comparison} shows the comparison between the Rayleigh--Plesset prediction (maroon curve) and the experimental data (blue circles) for all three cases. Good agreement is obtained over the full growth stage, confirming that the isentropic pressure law (equation~\ref{eq:Isentropic-compression}) adequately describes the cavitation bubble dynamics. The fitted values are: Case~1: $P_\text{o} = 0.95 \times 10^6$~Pa, $R_\text{o} = 5.8$~mm; Case~2: $P_\text{o} = 0.38 \times 10^6$~Pa, $R_\text{o} = 5.8$~mm; Case~3: $P_\text{o} = 0.35 \times 10^6$~Pa, $R_\text{o} = 9.3$~mm. These values establish the pressure scale of the cavitation event for each case and are used as inputs to the liquid column model in \S\ref{sec:liquid_column}.

\subsubsection{Dynamics of liquid column}\label{sec:liquid_column}
A growing cavitation bubble pushes the surrounding liquid outward, causing the liquid (liquid column) between the cavitation bubble and the air bubble to compress the air bubble. This compression increases the internal pressure of the air bubble, which then rebounds after reaching maximum compression. The rebound drives the liquid column outward, leading to its direct impact on the far boundary of the collapsing bubble. Throughout this process, the motion of the liquid column facilitates momentum exchange between the cavitation bubble and the air bubble. The following equation governs the motion of the liquid column driven by the pressure difference between the cavitation bubble and the air bubble,
\begin{equation}\label{eq:liquid_diff_eqn}
  \rho_\text{l}\, l_\text{lc}\,\frac{d^2 z}{dt^2} = P_\text{air bubble}(t) - P_\text{B}(t),
\end{equation}
Here, $\rho_\text{l}$ is the liquid density, $l_\text{lc}$ denotes the height of the liquid column, defined as ($h + d_\text{hole} - z_\text{org}$). $P_\text{air bubble}$ is the internal pressure of the air bubble, $P_\text{air bubble} = P_\text{atm}(z_\text{org}/z(t))^k$. $P_\text{atm}$ is atmospheric pressure, $z$ is the location of the air-water interface from the base of the blind hole at time $t$. $z_\text{org}$ is the uncompressed height of the air bubble, measured directly from experimental images. $P_\text{B}$ is the internal pressure in the cavitation bubble at time $t$, which is determined directly from equation~\ref{eq:Rayleigh_Plesset}. The model is developed under the following assumptions: (i) the liquid column moves as a rigid one-dimensional slug; and (ii) the column motion is driven solely by the pressure difference at its two ends, with no influence from the blind-hole walls. In this model, the pressure in the cavitation bubble drives the behaviour of the air bubble; however, the reverse effect is not considered, resulting in a one-way coupling. Equations~\ref{eq:Rayleigh_Plesset} and \ref{eq:liquid_diff_eqn} are solved simultaneously for each case, with the initial cavitation bubble pressure and radius ($P_0$, $R_0$) specified in \S\ref{sec:cav_model}. The model is solved numerically in MATLAB using the \texttt{ode45} solver.

\begin{figure}
\centering
\subfigure[]{\label{fig:cav_air_lc_case1}
\includegraphics[trim = 0mm 0mm 0mm 0mm, clip, angle=0,width=0.45\textwidth]{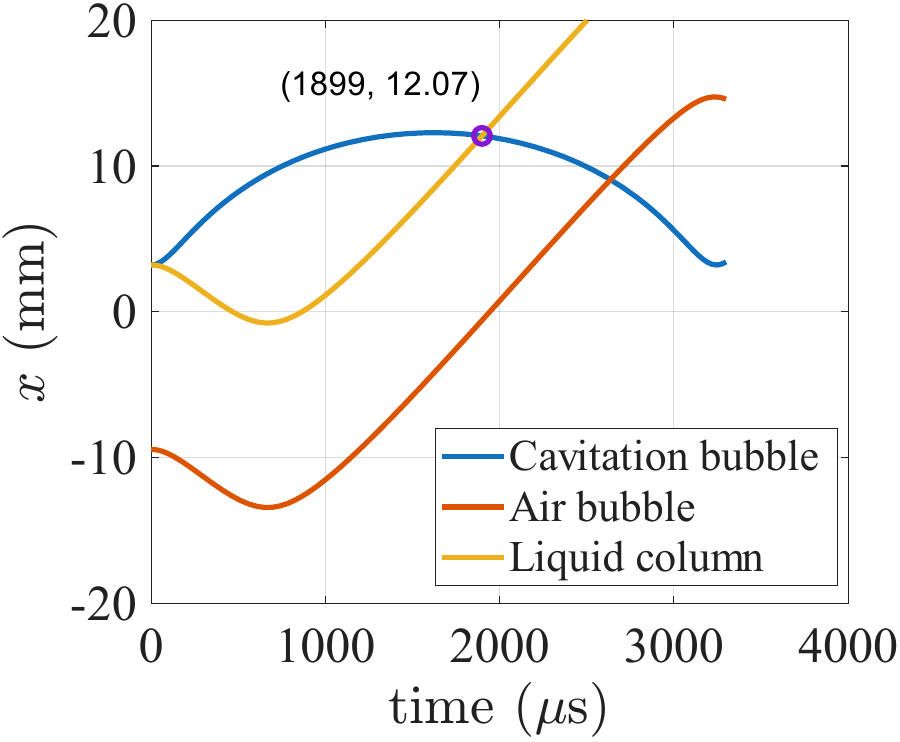}
}
\hspace{5mm}
\subfigure[]{\label{fig:cav_air_lc_case2}
\includegraphics[trim = 0mm 0mm 0mm 0mm, clip, angle=0,width=0.45\textwidth]{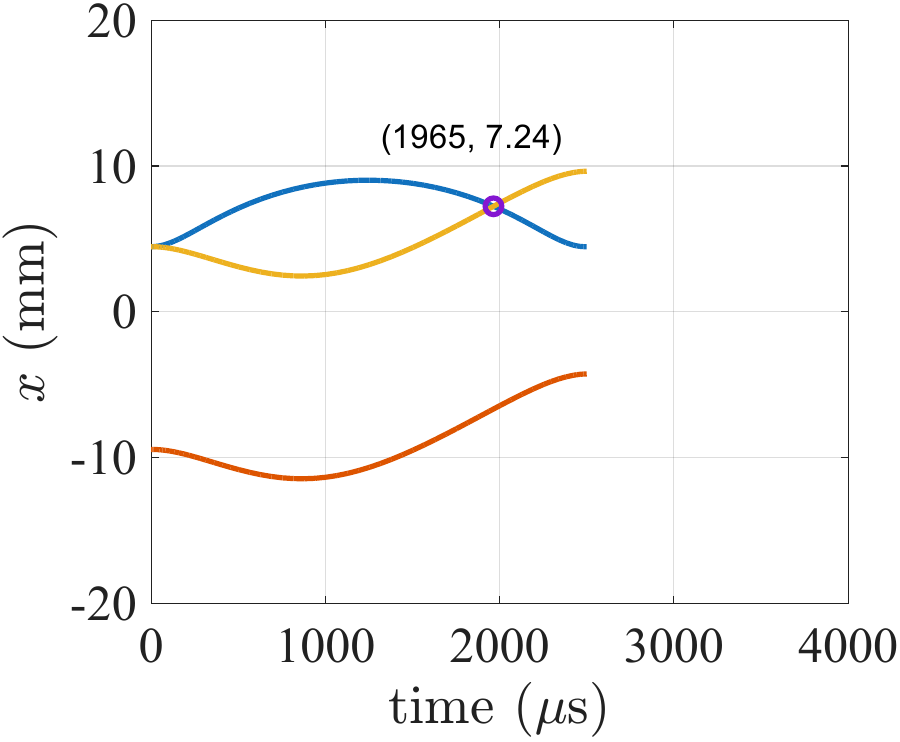}
}
\subfigure[]{\label{fig:cav_air_lc_case3}
\includegraphics[trim = 0mm 0mm 0mm 0mm, clip, angle=0,width=0.45\textwidth]{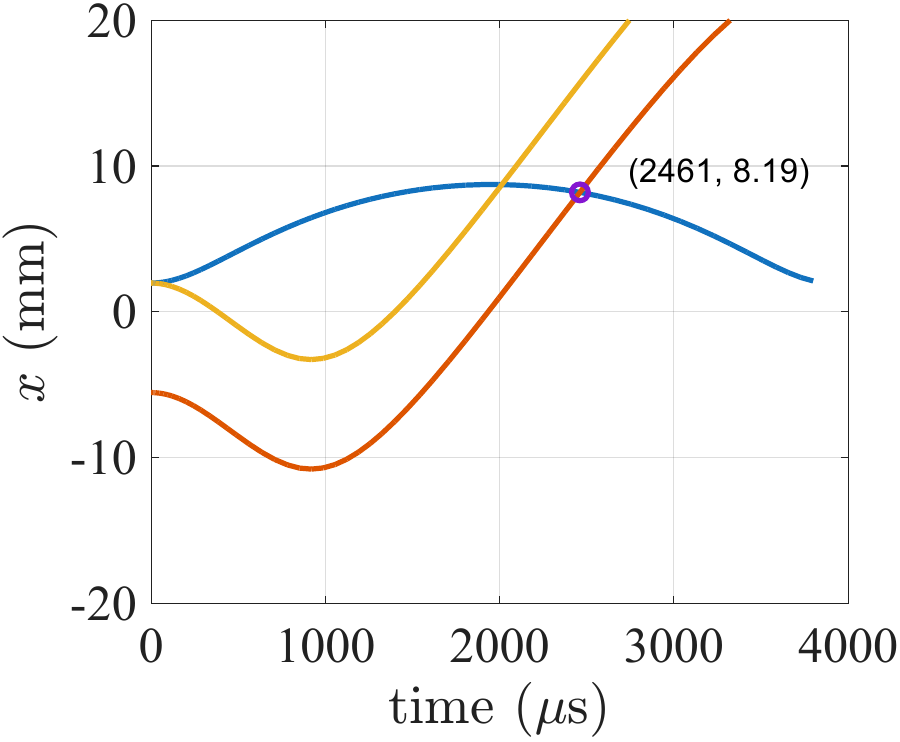}
}
\caption{Space--time trajectories of three boundaries for each case. Cyan curves: far boundary of the cavitation bubble during growth and collapse stages; Orange curves: air bubble upper boundary; Yellow circles: top of the liquid column (or penetrating bubble). A purple circle marks the predicted interaction point where the liquid column (or air bubble) meets the far cavitation boundary. (a)~Case~1 ($\mathcal{H} = 0.44$, $\mathcal{B} = 0.37$): liquid column impact at $t = 1900$~$\mu$s, $x = 12.07$~mm, during the collapse phase; (b)~Case~2 ($\mathcal{H} = 0.72$, $\mathcal{B} = 0.37$): late impact at $t = 1965$~$\mu$s, $x = 7.24$~mm, near end of collapse; (c)~Case~3 ($\mathcal{H} = 0.24$, $\mathcal{B} = 0.63$): air bubble boundary reaches far cavitation boundary at $t = 2461$~$\mu$s, $x = 8.19$~mm.}
\label{fig:cav_air_lc_trajectories}
\end{figure}
The initial values $P_0$ and $R_0$, determined from the Rayleigh--Plesset model by fitting the experimental growth stage of the cavitation bubble, ensure that the growth stage of the cavitation bubble in the liquid column model is consistent with the experiments. This consistency also leads to a similar cavitation bubble lifetime $t_\text{cav}$: experiments: 3300~$\mu$s, model: 3300~$\mu$s for Case~1; experiments: 2200~$\mu$s, model: 2500~$\mu$s for Case~2; and experiments: 3750~$\mu$s, model: 3800~$\mu$s for Case~3, as shown in Figure~\ref{fig:cav_air_lc_trajectories}. The trajectory of the air bubble-liquid column boundary (orange curve in Figure~\ref{fig:cav_air_lc_trajectories}) exhibits an initial compression followed by expansion through a point of maximum compression, in agreement with experiments. The liquid column (yellow curve) impacts the cavitation bubble trajectory (cyan curve) during the collapse stage.

The three space--time trajectories (cavitation bubble boundary, air bubble boundary, and liquid column top) are shown together in Figure~\ref{fig:cav_air_lc_trajectories}. A purple circle on each panel marks the predicted impact point. In Case~1, impact occurs at $t = 1900$~$\mu$s (experiments: 2800~$\mu$s) and $x = 12.07$~mm above the surface, well within the collapse stage ($t_\text{cav}/2 = 1650$~$\mu$s), confirming that the liquid column arrives during the collapse phase. In Case~2, the corresponding impact point ($t = 1965$~$\mu$s, $x = 7.24$~mm; experiments: 2175~$\mu$s) falls near the end of collapse, explaining the failure to form a ring. In Case~3, the liquid column and the air bubble boundary reach the cavitation boundary nearly at the same time and location ($t = 2461$~$\mu$s, $x = 8.19$~mm; experiments: 2650~$\mu$s). The model is developed under the assumption that the liquid column remains coherent, irrespective of the intensity of the air bubble rebound. In Case~3, however, this assumption breaks down, as the highly compressed air bubble passes through the liquid column (for large~$\mathcal{B}$) and impacts the boundary of the cavitation bubble.

\subsection{Dynamics of the air bubble}\label{sec:air_bubble}

\subsubsection{Compression mechanism}
\begin{figure}
\centering
\begin{minipage}{0.5\textwidth}
\centering
\captionsetup{
    width=\linewidth,
    font=small,
    justification=justified
}
\captionof{table}{Air bubble height at key stages for each representative case. 
$z_\text{org}$: initial (uncompressed) height; 
$z_\text{in}$: height at maximum compression, measured directly from the experimental images and used as the initial condition for the expansion model (equation~\ref{eq:air_diff_eqn}).}
\begin{tabular}{c c c c c c}
\hline
\textbf{Case} & 
\textbf{$\mathcal{B}$} & 
\textbf{$\mathcal{H}$} & 
\textbf{$z_\text{org}$ (mm)} & 
\textbf{$z_\text{in}$ (mm)} &
\textbf{$\Delta z$ (mm)}\\ \hline
Case 1 & 0.37 & 0.44 & 5.5 & 3.6 & 1.9 \\ \hline
Case 2 & 0.37 & 0.72 & 5.5 & 4.5 & 1 \\ \hline
Case 3 & 0.63 & 0.24 & 9.5 & 7.1 & 2.4 \\ \hline
\end{tabular}
\label{tab:Summary_air_bubble_th1}
\end{minipage}
\hfill
\begin{minipage}{0.4\textwidth}
\centering
\includegraphics[width=\textwidth]{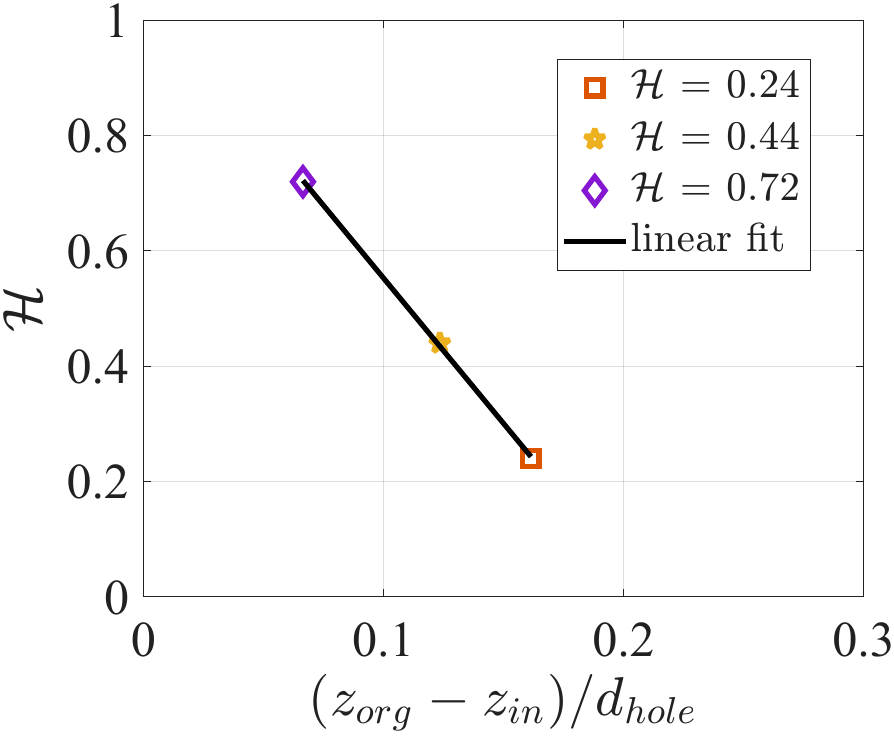}
\captionsetup{
    width=\linewidth,
    font=small,
    justification=justified
}
\caption{Variation of the height difference of the air bubble due to compression with $\mathcal{H}$. Here, the $x$-axis denotes the dimensionless height difference of the air bubble, normalized by the height of the blind hole, while the $y$-axis denotes $\mathcal{H}$. Data are shown for three key cases, along with a linear fit ($R^2=0.99$) with a slope of -5.04 and an intercept of 1.06.}
\label{fig:air_bubble_height}
\end{minipage}

\end{figure}

As the cavitation bubble grows above the blind hole, it drives a radially outward flow in all directions. The component of this flow directed downward into the hole pushes the liquid column inward, compressing the air bubble at the base. The compression is quantified as $\Delta z = z_\text{org} - z_\text{in}$ (Table~\ref{tab:Summary_air_bubble_th1}), where $z_\text{org}$ is the uncompressed height and $z_\text{in}$ is the height at maximum compression, both measured from the experimental images. As shown in Table~\ref{tab:Summary_air_bubble_th1}, $\Delta z$ increases systematically with decreasing $\mathcal{H}$ (see Figure~\ref{fig:air_bubble_height}). When the cavitation bubble is generated closer to the surface, its flow field induces a stronger downward velocity at the hole entrance, resulting in greater compression.

\subsubsection{Expansion model and liquid column speed}
\begin{figure}
\centering
\subfigure[]{\label{fig:air_model_case1}
\includegraphics[trim = 0mm 0mm 0mm 0mm, clip, angle=0,width=0.45\textwidth]{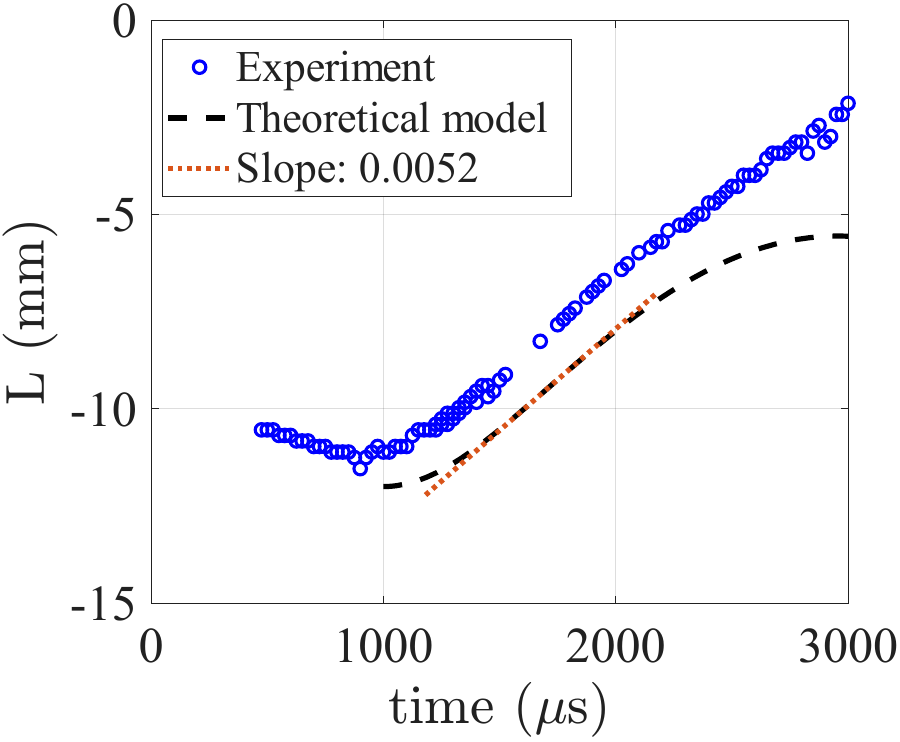}
}
\hspace{5mm}
\subfigure[]{\label{fig:air_model_case2}
\includegraphics[trim = 0mm 0mm 0mm 0mm, clip, angle=0,width=0.45\textwidth]{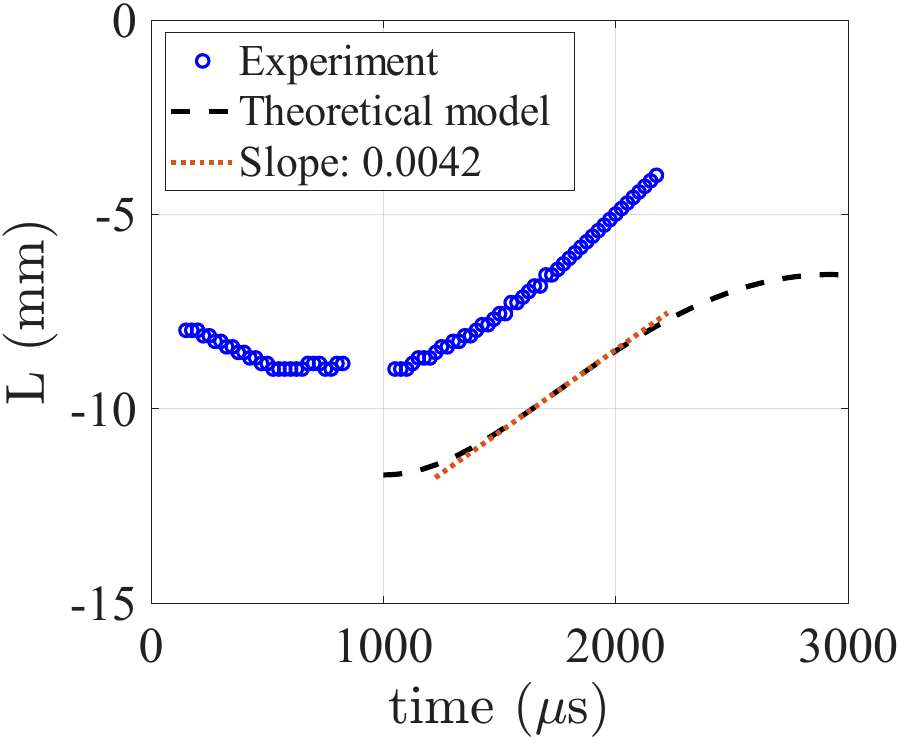}
}
\subfigure[]{\label{fig:air_model_case3}
\includegraphics[trim = 0mm 0mm 0mm 0mm, clip, angle=0,width=0.45\textwidth]{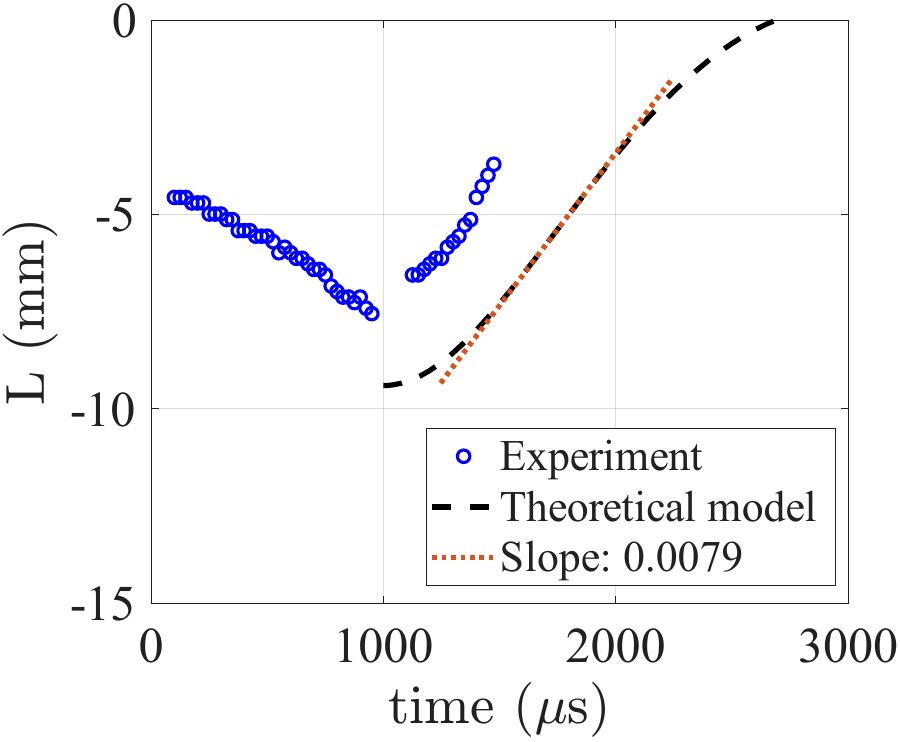}
}
\caption{Air bubble boundary position $L = z - d_\text{hole}$ as a function of time. Blue circles: experimental data (compression then expansion). Black dashed curve: prediction of equation~\ref{eq:air_diff_eqn}, applied from the point of maximum compression ($z = z_\text{in}$). Red dotted line: maximum slope of the model curve, from which the terminal liquid column speed $U_\text{lc}$ is extracted. Negative $L$ indicates the interface lies inside the blind hole. (a)~Case~1 ($\mathcal{H} = 0.44$, $\mathcal{B} = 0.37$), $z_\text{tuned}=3.3$~mm, $U_\text{lc} = 5.2$~m/s; (b)~Case~2 ($\mathcal{H} = 0.72$, $\mathcal{B} = 0.37$), $z_\text{tuned}=3.9$~mm, $U_\text{lc} = 4.2$~m/s; (c)~Case~3 ($\mathcal{H} = 0.24$, $\mathcal{B} = 0.63$), $z_\text{tuned}=5.6$~mm, $U_\text{lc} = 7.9$~m/s.}
\label{fig:air_bubble_expansion_modeling}
\end{figure}

Once compressed to the minimum height $z_\text{in}$, the air bubble achieves a maximum internal pressure $P_\text{bmax}$, estimated from isentropic compression:
\begin{equation}
P_\text{bmax} = P_\text{atm}\left(\frac{A_\text{hole} z_\text{org}}{A_\text{hole} z_\text{in}}\right)^k,
\end{equation}
where $P_\text{atm}$ is the atmospheric pressure and $A_\text{hole}$ is the area of the cross section of the blind hole. The pressurised air bubble then expands vertically (radially confined), driving the liquid column above it upward and out of the hole. The equation of motion for the air--water interface position $z(t)$, measured from the base of the blind hole, is:
\begin{equation}\label{eq:air_diff_eqn}
  \rho_\text{l}\, l\,\frac{d^2 z}{dt^2} = P_\text{bmax}\left(\frac{z_\text{in}}{z}\right)^k - P_\text{atm},
\end{equation}
where $l = d_\text{hole} - z_\text{org}$ is the initial height of the liquid column above the air bubble inside the blind hole. The model assumes: (i) one-dimensional vertical motion; (ii) the ambient pressure above the liquid column remains atmospheric throughout (this assumption breaks down when the cavitation bubble begins to collapse, which accounts for the deviations seen at $t \gtrsim 2000~\mu$s in Figure~\ref{fig:air_bubble_expansion_modeling}). Equation~\ref{eq:air_diff_eqn} is integrated numerically using MATLAB's \texttt{ode45} solver, with $z_\text{in,th}$ ($z_\text{in,th} = z_\text{tuned}$) determined iteratively to match the trajectory of the expanding air bubble.

Figure~\ref{fig:air_bubble_expansion_modeling} shows the air bubble boundary position $L = z - d_\text{hole}$ as a function of time for all three cases. The experimental data (blue circles) first decrease (compression phase) and then increase through a minimum (expansion phase); $L < 0$ indicates the interface is inside the blind hole. The model prediction (black dashed curve) is applied only during the expansion phase, starting from $z_\text{tuned}$. For Case~1 (Figure~\ref{fig:air_model_case1}), model and experiment agree well for $t \lesssim 2000~\mu$s, after which the growing cavitation bubble modifies the local pressure field and the atmospheric-pressure assumption breaks down. The maximum slope of the predicted trajectory (red dotted line) gives the terminal liquid column speed: $U_\text{lc} = 5.2$~m/s for Case~1, $4.2$~m/s for Case~2, and $7.9$~m/s for Case~3. Once launched at this speed, the liquid column travels under inertia and maintains approximately the same speed until impact.

The three speeds directly reflect the three distinct outcomes. In Case~1 ($U_\text{lc} = 5.2$~m/s), the liquid column travels far enough and fast enough to impact the far cavitation boundary during the collapse phase, producing the vortex ring. In Case~2 ($U_\text{lc} = 4.2$~m/s), the slower column arrives only near the very end of collapse, after the collapsing jet has formed, and no ring results. In Case~3 ($U_\text{lc} = 7.9$~m/s), the model overpredicts the actual column speed because the rapidly expanding air bubble overtakes and fragments the short liquid column ($\mathcal{B} = 0.63$) before it can be treated as a coherent slug; the air bubble then impacts the cavitation boundary directly.

\subsubsection{A timing criterion for vortex ring formation}\label{sec:Pi}
The analysis above suggests a simple criterion for vortex ring formation: the liquid column must travel from the hole entrance to the far cavitation boundary during the collapse phase of the cavitation bubble. The relevant dimensionless parameter is the ratio of the liquid column travel time to the half-lifetime of the cavitation bubble:
\begin{equation}\label{eq:Pi}
  \Pi = \frac{ h + R_{\max}}{U_\text{lc} \cdot (t_\text{cav}/2)},
\end{equation}
where $h$ is the location of the cavitation bubble generation from solid surface and $R_{\max}$ is the maximum cavitation bubble radius ($h + R_\text{max}$ represents the travel distance from the hole entrance to the far boundary). $U_\text{lc}$ is the liquid column speed from the expansion model and $t_\text{cav}/2$ is the half-lifetime of the cavitation bubble (the collapse timescale). The parameter $\Pi$ compares these two timescales: if $\Pi < 1$, the liquid column would arrive at the far boundary before the midpoint of the bubble lifetime, i.e. during the growth phase, which is too early. If $\Pi > 1$, the liquid column takes longer than a half-lifetime to arrive, meaning it either impacts the boundary during the collapse phase (moderate $\Pi$) or arrives after complete collapse (large $\Pi$).

Table~\ref{tab:Pi_values} shows the computed values of $\Pi$ for the three representative cases. 
\begin{table}
\centering
\caption{Timing parameter $\Pi = (h + R_\text{max})/(U_\text{lc} \cdot t_\text{cav}/2)$ for each representative case. $U_\text{lc}$ is the terminal liquid column speed from the expansion model; $t_\text{cav}$ is the total bubble lifetime from Table~\ref{tab1:Summary}. Ring formation is observed for $1 \lesssim \Pi \lesssim 1.5$.}
\begin{tabular}{c c c c c c c c}
\hline
\textbf{Case} & \textbf{$\mathcal{H}$} & \textbf{$\mathcal{B}$} & \textbf{$R_{\max}$ (mm)} & \textbf{$U_\text{lc}$ (m/s)} & 
\textbf{$\Pi$} \\ \hline
Case 1 & 0.44 & 0.37 & 7.3 & 5.2 & 1.22 \\
Case 2 & 0.72 & 0.37 & 6.2 & 4.2 & 2.30  \\
Case 3 & 0.24 & 0.63 & 8.2 & 7.9 & 0.69 \\ \hline
\end{tabular}
\label{tab:Pi_values}
\end{table}
For Case~1 ($\Pi = 1.22$), the liquid column arrives at approximately $1.22 \times (t_\text{cav}/2) = 2025$~$\mu$s after the start, which falls just after the midpoint of the bubble lifetime ($t_\text{cav}/2 = 1650$~$\mu$s). This places the impact squarely within the collapse phase, consistent with the observation of a vortex ring. For Case~2 ($\Pi = 2.30$), the travel time exceeds the half-lifetime, so the liquid column arrives well into the collapse stage, near the end of collapse, and finds a jet-dominated flow rather than a collapsing boundary amenable to ring formation. For Case~3 ($\Pi = 0.69$), the computed travel time is shorter than the half-lifetime, implying that the column would arrive during the growth phase. However, accounting for the column height gives $\Pi_\mathcal{B} = (l + h + R_\text{max})/[U_\text{lc}(t_\text{cav}/2)] = 1.07$, indicating that the air–water interface reaches the far boundary just after the midpoint of the bubble lifetime. As discussed in \S\ref{sec:air_bubble}, the model overestimates $U_\text{lc}$ in this case because the air bubble bypasses the short liquid column; the column never reaches the far boundary as a coherent slug at all.

These results delineate two distinct failure modes with respect to the formation of vortex rings. The \emph{late-arrival} failure (Case~2, large $\Pi$) occurs when $\mathcal{H}$ is too large: the air bubble is compressed less, expands more slowly, and the liquid column speed is insufficient to reach the far boundary before the bubble collapses. The \emph{bypass} failure (Case~3) occurs when $\mathcal{B}$ is too large: the liquid column is too short to remain ahead of the rapidly expanding air bubble. Vortex ring formation is observed when both conditions are avoided, corresponding to the regime $\mathcal{H} \lesssim 0.5$ and $\mathcal{B} \lesssim 0.5$ identified in Figure~\ref{fig:All_data}. Within this regime, $\Pi$ takes an intermediate value ($1 \lesssim \Pi \lesssim 1.5$) indicating that the liquid column arrives during the collapse phase as a coherent slug.

\subsection{Vortex ring: formation, behavior, and trajectory}\label{sec:ring_trajectory}
\begin{figure}
\centering
\includegraphics[trim = 0mm 0mm 0mm 0mm, clip, angle=0,width=0.7\textwidth]{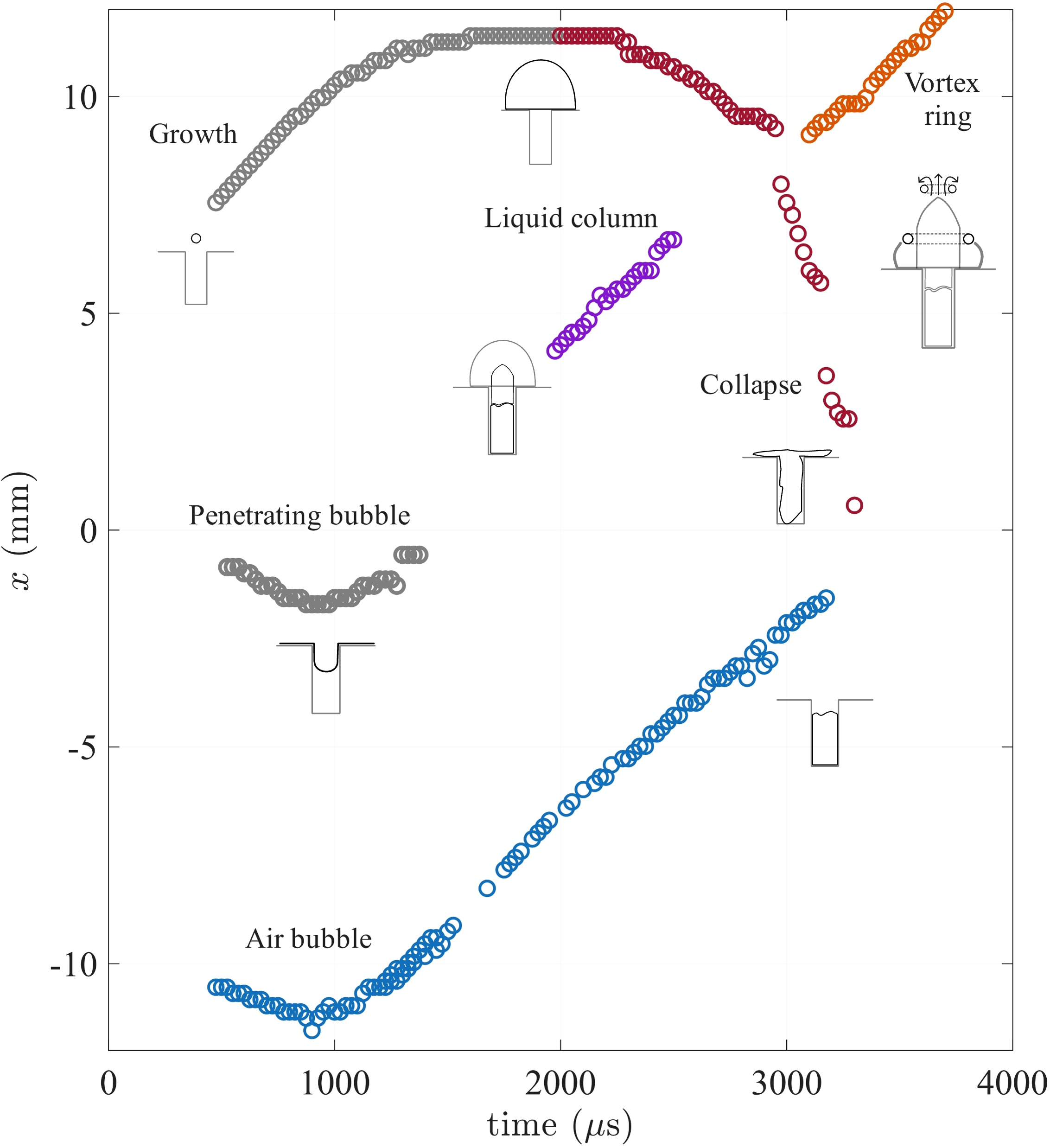}
\caption{Space--time map for Case~1 ($\mathcal{H} = 0.44$, $\mathcal{B} = 0.37$), showing all observed processes on a single plot. The $y$-axis is distance from the top surface ($x$) of the block (negative values are inside the blind hole); $t = 0$ is bubble generation. Grey circles: far boundary of the cavitation bubble during growth; maroon circles: during collapse; cyan circles: air bubble upper boundary; grey circles: penetrating bubble boundary; purple circles: liquid column top; orange circles: vortex ring centre. The kink in the maroon (collapse) trajectory marks the liquid column impact, immediately followed by ring formation (orange circles). Schematics at selected time stamps illustrate the physical state of the system.}
\label{fig:Spatial_regime_map_case1}
\end{figure}

The vortex ring emerges as a result of the interaction between the cavitation bubble and the confined air bubble located in the blind hole. Figure \ref{fig:All_data} shows several experimental data points indicated by orange circles where the vortex ring forms. These circles fall within a region of the parameter space $\mathcal{B}$-$\mathcal{H}$ characterised by $\mathcal{H} \lesssim 0.5$ and $\mathcal{B} \lesssim 0.5$.
The corresponding representative case discussed in the present study, Case~1 (discussed in \S \ref{sec:case-1}), describes the processes which lead to the formation of the vortex ring at $\mathcal{H} =0.44$ and $\mathcal{B} = 0.37$. Figure~\ref{fig:Spatial_regime_map_case1} illustrates a space-time map for Case~1 that represents all processes that occurred during the growth and collapse stages of the cavitation bubble. This map offers an advantage over the process map (Figure~\ref{fig:case-1_phenomena}) and the cavitation time map (Figure~\ref{fig:Time_Stamp}), as it visually shows the processes and their mutual interactions. To make the map more informative, schematics are attached to each curve, representing the processes at specific time stamps. The top-center of the blind hole is considered the origin of the coordinate system in the map in Figure~\ref{fig:Spatial_regime_map_case1}. The processes are shown on both sides of the $x$-axis. This indicates that the processes depicted below the $x$-axis occurred within the blind hole. The start and end points of the entire event, i.e., the growth and collapse stages of the cavitation bubble, are represented by the curves marked with grey and maroon circles, respectively. These circles denote the location of the uppermost point on the boundary of the cavitation bubble. The trajectory of the growth stage is compared with the Rayleigh--Plesset model and results in good agreement, as shown in Figure~\ref{fig:rp_comparison}.

As discussed in \S\ref{sec:case-1}, three key processes occur during the entire event: (i) air bubble compression and expansion, (ii) penetrating bubble entry and exit, and (iii) motion of the liquid column. The onset of bubble compression (cyan circles) and penetrating bubble entry (grey circles) occur nearly simultaneously, as do bubble expansion and the onset of penetrating bubble exit. Figure~\ref{fig:Spatial_regime_map_case1} clearly shows that the complete oscillation of the air bubble and the penetrating bubble occurred during the growth stage of the cavitation bubble. The air bubble, after reaching its original height, continued to expand within the blind hole, as indicated by the cyan circles. The grey circles of the penetrating bubble and the cyan circles of the air bubble are shown below the $x$-axis, indicating that both processes occurred within the blind hole. The height difference between the trajectories of the air bubble and the penetrating bubble denotes the liquid column. The cyan circles are approaching the $x$-axis, indicating that the air bubble boundary is approaching the top end of the blind hole and thereby pushing the liquid column. The purple circles represent the other end of the liquid, i.e., the top end of the liquid column. 
Some purple circles between the end of the penetrating bubble and the motion of the liquid column are not shown due to obstruction caused by luminescence and metal fumes (black patches) captured in the images. Similarly, the motion of the liquid column could not be captured just before impact. The trajectory of the liquid column (purple circles) shows a noticeably higher slope than that of the expanding air bubble (cyan circles). This occurs due to the high initial momentum imparted to the liquid column by the expanding air bubble, while the air bubble decelerates as it undergoes excessive expansion. If the trajectory of the liquid column is extrapolated to the collapse stage, it intersects the location of the kink. This kink, shown in the trajectory of the cavitation collapse (maroon circles), marks the impact with the liquid column and the subsequent formation of the vortex ring, as indicated by the orange circles.
%
\begin{figure}
\centering
\subfigure[]{\label{fig:vortex_ring_trajectory}
\includegraphics[trim = 0mm 0mm 0mm 0mm, clip, angle=0,width=0.4\textwidth]{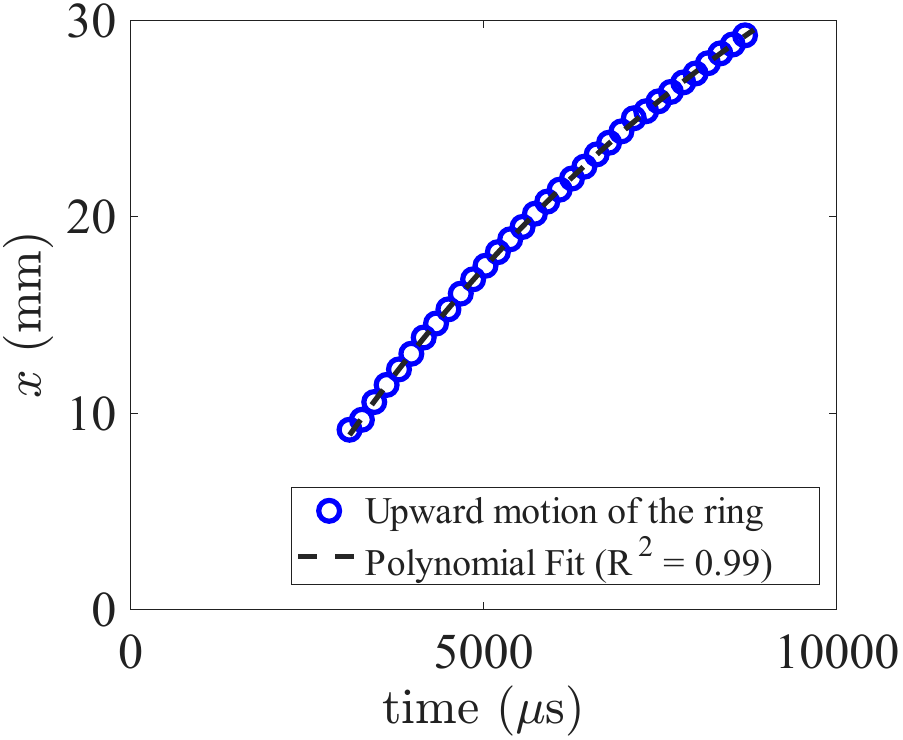}
}
\hspace{5mm}
\subfigure[]{\label{fig:vortex_ring_speed}
\includegraphics[trim = 0mm 0mm 0mm 0mm, clip, angle=0,width=0.4\textwidth]{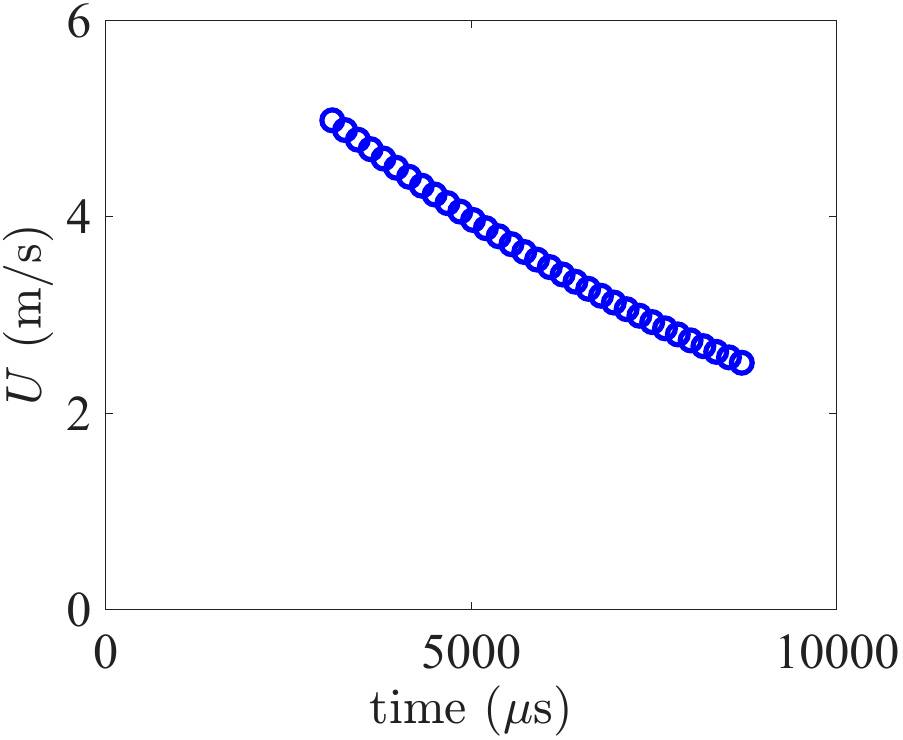}
}
\caption{Vortex-ring characteristics: (a) Variation of the vertical distance traveled by the vortex ring with time. Here, $x$ represents the distance from the solid surface. Blue circles denote the center positions of the vortex ring, and the dashed curve represents a third-degree polynomial fit with $R^2 = 0.99$. (b) Variation of the vortex ring speed ($U$) with time. The speed exhibits a continuous quadratic decay, decreasing from 5 m/s to 2.5 m/s over a few milliseconds.}
\end{figure}
The ring forms immediately after the impact of the liquid column on the far boundary of the cavitation bubble. This mechanism is distinct from the vortex ring reported by \cite{Lew2007} near a through-hole, where the ring was attributed to circulation induced by the jet passing across the hole. Here, the ring arises from the impulsive impact of a directed liquid slug on a curved, collapsing interface, a mechanism enabled specifically by the confinement of the air bubble in the blind hole.
The vortex ring was tracked and quantitative data was extracted using image processing techniques. Figure~\ref{fig:Ring-Track} shows the trajectory of the vortex ring, obtained by combining frames extracted from the image stack at fixed 500~$\mu$s intervals between 3225~$\mu$s and 10000~$\mu$s.

The vertical distance traversed by the ring as a function of time is fitted with a third-degree polynomial (see Figure~\ref{fig:vortex_ring_trajectory}), and the ring speed decreases quadratically with time, as shown in Figure~\ref{fig:vortex_ring_speed}. In Case~1, the speed of the vortex ring decreases from an initial value of $U = 5$~m/s to $2.5$~m/s in just a few milliseconds. 
The vortex ring moves upward with a Reynolds number of $Re \approx 4500$, where $Re = Ua/\nu$, $U$ denotes the speed of the vortex ring, $a$ denotes the diameter of the ring, and $\nu$ denotes the kinematic viscosity of the surrounding medium. The Reynolds number of $Re \approx 4500$ ($a=0.9$~mm and $\nu = 1\times 10^{-6}$~m$^2$/s) is considered high, which causes the vortex ring to become unstable and disintegrate (\cite{maxworthy1972structure}). The composite image in Figure~\ref{fig:Ring-Track} shows the ring breaking apart as it rises. The shedding of vorticity, azimuthal instabilities, and the transition to turbulence are likely the primary causes that make the ring unstable. 
\section{\label{sec:discussion}Conclusions}
\begin{figure}
\centering
\includegraphics[trim = 0mm 0mm 0mm 0mm, clip, angle=0,width=0.75\textwidth]{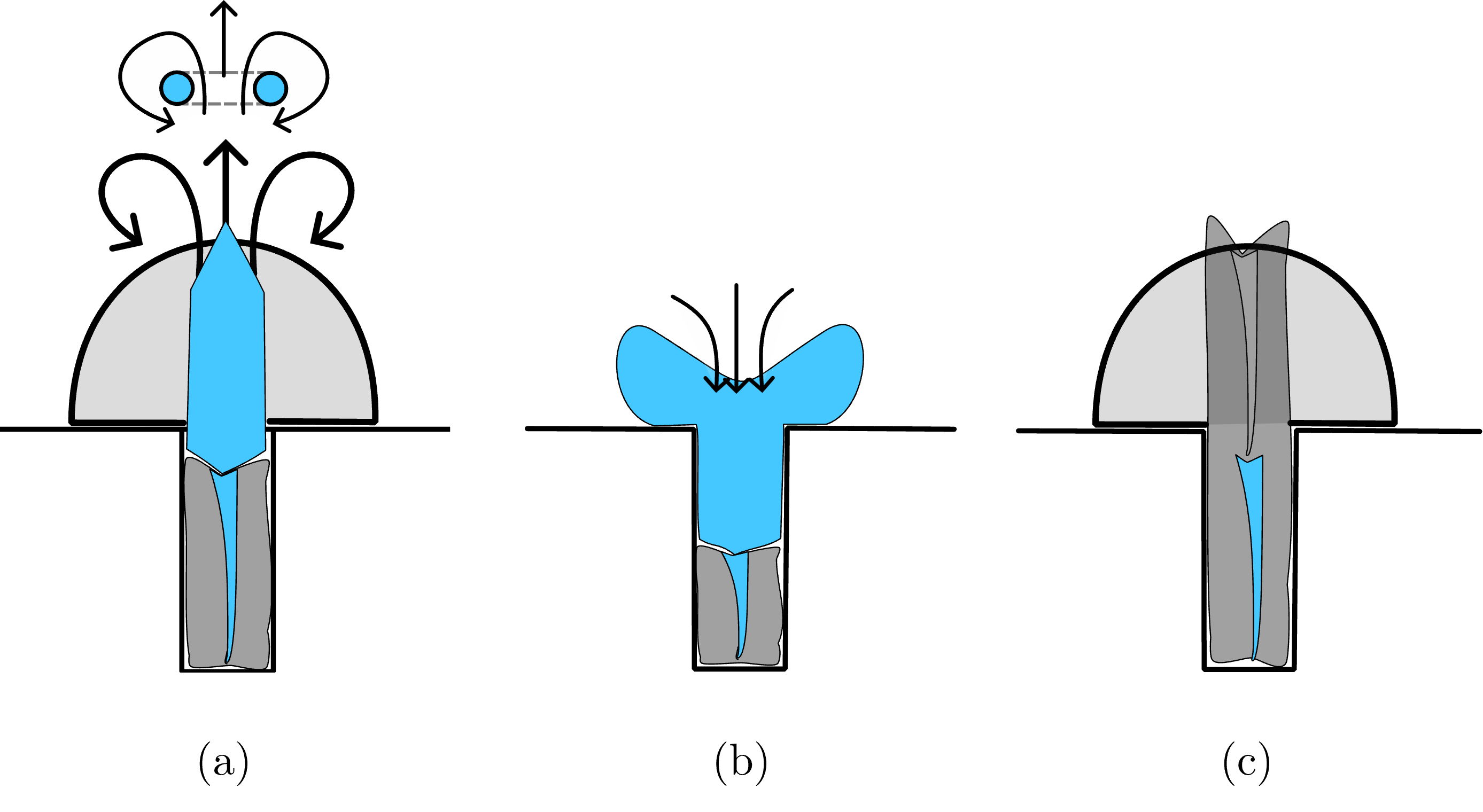}
\caption{Schematic summarizing the three interaction outcomes between a growing cavitation bubble and a confined air bubble. Light grey hemisphere: cavitation bubble at maximum growth. Dark grey cylinder: expanding air bubble. Light blue: liquid column. (a)~Case~1 ($\mathcal{H} = 0.44$, $\mathcal{B} = 0.37$): the liquid column (L-B impact) strikes the far cavitation boundary during the collapse phase, producing a vortex ring. (b)~Case~2 ($\mathcal{H} = 0.72$, $\mathcal{B} = 0.37$): the liquid column exits the hole only near the very end of collapse (Late Impact, black arrows denote the collapsing jet), producing no ring. (c)~Case~3 ($\mathcal{H} = 0.24$, $\mathcal{B} = 0.63$): the expanding air bubble bypasses the short liquid column and directly impacts the far boundary (B-B impact), producing no ring.}
\label{fig:Cartoon}
\end{figure}

This paper has reported an experimental and theoretical study of vortex ring formation arising from the interaction between a spark-generated cavitation bubble and a confined air bubble in a blind hole. The main conclusions are as follows.

\begin{enumerate}[(i)]
  \item \emph{Three distinct interaction outcomes}: Depending on the dimensionless stand-off distance $\mathcal{H} = h/R_{\max}$ and the air bubble fill fraction $\mathcal{B} = (d_\text{hole} - d_\text{top})/d_\text{hole}$, three qualitatively different outcomes are observed across the parameter space $\mathcal{H} < 1$, $\mathcal{B} < 1$ (Figure~\ref{fig:All_data}): (a)~liquid column impact on the far cavitation boundary during the collapse phase (L-B Impact), leading to vortex ring formation; (b)~late liquid column impact near the end of collapse (Late Impact), producing no ring; and (c)~the expanding air bubble directly striking the far boundary after bypassing the liquid column (B-B Impact), also producing no ring. These three outcomes are summarised schematically in Figure~\ref{fig:Cartoon}.
  \item \emph{Vortex ring formation mechanism}: When $\mathcal{H} \lesssim 0.5$ and $\mathcal{B} \lesssim 0.5$, the growing cavitation bubble drives a downward flow into the blind hole, compressing the confined air bubble. The compressed air bubble then re-expands, driving the liquid column above it upward as a coherent slug. The slug impacts the far boundary of the cavitation bubble during its collapse phase, generating an axisymmetric impulse that creates a sheet of azimuthal vorticity which rolls up and detaches as a vortex ring. This mechanism requires that the liquid column remain intact ahead of the expanding air bubble (condition on $\mathcal{B}$) and that it arrive during the collapse phase rather than after it (condition on $\mathcal{H}$).
  \item \emph{Role of $\mathcal{H}$ -- compression and timing}: The air bubble compression increases linearly as $\mathcal{H}$ decreases (Figure~\ref{fig:air_bubble_height}). Smaller $\mathcal{H}$ produces a closer cavitation bubble, which induces a stronger downward flow into the hole, resulting in greater compression and higher $P_\text{bmax}$. This leads to faster air bubble expansion and higher liquid column speed $U_\text{lc}$. When $\mathcal{H} \gtrsim 0.5$, the expansion is too slow and the liquid column arrives too late, constituting the Late Impact failure mode. The penetrating bubble (the portion of the cavitation bubble that enters the blind hole when $\mathcal{H} \lesssim 0.5$) exits the hole and forms a conical tip on the liquid column, augmenting its effective height and facilitating timely impact.
  \item \emph{Role of $\mathcal{B}$ -- liquid column coherence}: The air bubble fill fraction $\mathcal{B}$ controls the height of the liquid column above the air bubble. When $\mathcal{B} \gtrsim 0.5$, the column is too short for the rapidly expanding air bubble to push coherently; the air bubble overtakes and fragments the column, and then impacts the cavitation boundary directly (B-B impact). No vortex ring results from this direct contact.
  \item \emph{Timing criterion $\Pi$}: The condition for vortex ring formation is captured quantitatively by the dimensionless parameter $\Pi = (h + R_{\max})/(U_\text{lc} \cdot t_\text{cav}/2)$ (equation~\ref{eq:Pi}), which compares the liquid column travel time to the cavitation collapse timescale. For the three representative cases, $\Pi = 1.22$ (Case~1, ring forms), $\Pi = 2.30$ (Case~2, late impact, no ring), and $\Pi = 0.69$ (Case~3, bypass, no ring, $\Pi_\mathcal{B} = 1.07$). These values suggest that vortex ring formation requires $1 \lesssim \Pi \lesssim 1.5$, corresponding to liquid column impact during the collapse phase.
  \item \emph{Vortex ring trajectory}: The ring propagates away from the blind hole with an initial speed of $U \approx 5$~m/s, decelerating quadratically to $2.5$~m/s within a few milliseconds. The ring travels with a Reynolds number $Re = Ua/\nu \approx 4500$ ($a = 0.9$~mm), which lies above the threshold for azimuthal instability reported by \cite{maxworthy1972structure}, and the ring is observed to break apart as it rises.
\end{enumerate}
The connection between experiments and theory is made with the help of two simple one-dimensional models. The \emph{liquid column model} requires the cavitation bubble growth trajectory extracted from experimental images to predict the impact point, while the \emph{air bubble expansion model} requires experimental inputs about air bubble expansion trajectory. These models neglect interfacial instabilities, effects of the confining walls in the blind hole, compressibility of the liquid slug, and complex relationship between pressure and volume of the air bubble. Despite these simplifications, the models accurately capture essential features of cavitation-air bubble interaction, and provide a simple timing criterion for predicting the formation of vortex rings. Incorporating some of these complexities are left for a future study. Furthermore, the present experiments were restricted to a blind hole aspect ratio of three. Blind holes with higher aspect ratio, i.e. taller but narrow holes, can potentially suppress interfacial instabilities observed in the present study.

\vspace{5mm}
\backsection[Acknowledgements]{We wish to acknowledge the support of the Science and Engineering Research Board (SERB), Dept. of Science and Technology, India for funding this research through grant no. CRG/2021/007096, SR-FST-ETI-397-2015 and CRG/2021/004102. We acknowledge G. Prashanth Kumar (GPK) for conducting the initial experiments for this study.}

\backsection[Declaration of AI tools]{We acknowledge the use of following generative AI tools -- ChatGPT and Claude AI. The use of these tools was limited to polishing some of the text in the manuscript. No AI tool was used to generate new text or figure for the manuscript.}


\backsection[Declaration of interests]{The authors report no conflict of interest.}

\backsection[Data availability statement]{Videos of the three cases and a brief description are available at \url{https://people.iith.ac.in/hdixit/Cavitation_supplementary.html}. See JFM's \href{https://www.cambridge.org/core/journals/journal-of-fluid-mechanics/information/journal-policies/research-transparency}{research transparency policy} for more information}

\backsection[Author ORCIDs]{ Authors may include the ORCID identifiers as follows: 
Charul Gupta, \href{https://orcid.org/0009-0003-5518-2999}{https://orcid.org/0009-0003-5518-2999}; 
Yashwant Singh, \href{https://orcid.org/0009-0001-1099-5975}{https://orcid.org/0009-0001-1099-5975}; 
Lakshmana D Chandrala, \href{https://orcid.org/0000-0002-2695-9469}{https://orcid.org/0000-0002-2695-9469}; 
Harish N Dixit, \href{https://orcid.org/0000-0003-2993-7633}{https://orcid.org/0000-0003-2993-7633}; 
Badarinath Karri, \href{https://orcid.org/0000-0003-3958-1212}{https://orcid.org/0000-0003-3958-1212}.}

\backsection[Author contributions]{HND and BK conceptualized and supervised the research. CG and YS performed most of the experiments in this study. CG analysed the data with help from LDC, BK and HND. CG, LDC, BK and HND wrote the manuscript.}

\bibliographystyle{jfm}
\bibliography{cavitation}

@article{rayleigh1917,
  title={VIII. On the pressure developed in a liquid during the collapse of a spherical cavity},
  author={Rayleigh, Lord},
  journal={The London, Edinburgh, and Dublin Philosophical Magazine and Journal of Science},
  volume={34},
  number={200},
  pages={94--98},
  year={1917},
  publisher={Taylor \& Francis}
}

@article{benjamin1966collapse,
  title={The collapse of cavitation bubbles and the pressures thereby produced against solid boundaries},
  author={Benjamin, T Brooke and Ellis, Ao T},
  journal={Philosophical Transactions for the Royal Society of London. Series A, Mathematical and Physical Sciences},
  pages={221--240},
  year={1966},
  publisher={JSTOR}
}

@book{brennen2014cavitation,
  title={Cavitation and bubble dynamics},
  author={Brennen, Christopher E},
  year={2014},
  publisher={Cambridge university press}
}

@article{lauterborn1975,
  title={Experimental investigations of cavitation-bubble collapse in the neighbourhood of a solid boundary},
  author={Lauterborn, W and Bolle, H},
  journal={Journal of Fluid Mechanics},
  volume={72},
  number={2},
  pages={391--399},
  year={1975},
  publisher={Cambridge University Press}
}

@article{zhang1993,
  title={The final stage of the collapse of a cavitation bubble near a rigid wall},
  author={Zhang, Sheguang and Duncan, James H and Chahine, Georges L},
  journal={Journal of Fluid Mechanics},
  volume={257},
  pages={147--181},
  year={1993},
  publisher={Cambridge University Press}
}

@article{blake1981,
  title={Growth and collapse of a vapour cavity near a free surface},
  author={Blake, John R and Gibson, DC},
  journal={Journal of Fluid Mechanics},
  volume={111},
  pages={123--140},
  year={1981},
  publisher={Cambridge University Press}
}

@article{blake1987cavitation,
  title={Cavitation bubbles near boundaries},
  author={Blake, John R and Gibson, DC},
  journal={Annual review of fluid mechanics},
  volume={19},
  number={1},
  pages={99--123},
  year={1987},
  publisher={Annual Reviews 4139 El Camino Way, PO Box 10139, Palo Alto, CA 94303-0139, USA}
}

@article{plesset1971,
  title={Collapse of an initially spherical vapour cavity in the neighbourhood of a solid boundary},
  author={Plesset, Milton S and Chapman, Richard B},
  journal={Journal of Fluid Mechanics},
  volume={47},
  number={2},
  pages={283--290},
  year={1971},
  publisher={Cambridge University Press}
}

@article{Kauer2018,
author = {Kauer, Markus and Belova-Magri, Valentina and Cair{\'{o}}s, Carlos and Linka, Gerd and Mettin, Robert},
doi = {10.1016/j.ultsonch.2018.04.015},
issn = {18732828},
journal = {Ultrasonics Sonochemistry},
number = {April},
pages = {39--50},
pmid = {30080564},
publisher = {Elsevier},
title = {{High-speed imaging of ultrasound driven cavitation bubbles in blind and through holes}},
url = {https://doi.org/10.1016/j.ultsonch.2018.04.015},
volume = {48},
year = {2018}
}

@article{Bai2011,
author = {Bai, Li Xin and Xu, Wei Lin and Li, Chao and Gao, Yan Dong},
doi = {10.1016/S1001-6058(10)60150-3},
issn = {10016058},
journal = {Journal of Hydrodynamics},
number = {5},
pages = {562--569},
publisher = {Publishing House for Journal of Hydrodynamics},
title = {{The counter-jet formation in an air bubble induced by the impact of shock waves}},
url = {http://dx.doi.org/10.1016/S1001-6058(10)60150-3},
volume = {23},
year = {2011}
}

@article{Karri2012,
author = {Karri, Badarinath and Avila, Silvestre Roberto Gonzalez and Loke, Yee Chong and O'Shea, Sean J. and Klaseboer, Evert and Khoo, Boo Cheong and Ohl, Claus Dieter},
doi = {10.1103/PhysRevE.85.015303},
issn = {15393755},
journal = {Physical Review E - Statistical, Nonlinear, and Soft Matter Physics},
title = {{High-speed jetting and spray formation from bubble collapse}},
year = {2012}
}

@article{GonzalezAvila2015,
author = {{Gonzalez Avila}, Silvestreroberto R. and Song, Chaolong and Ohl, Claus Dieter},
doi = {10.1017/jfm.2015.33},
issn = {14697645},
journal = {Journal of Fluid Mechanics},
pages = {31--51},
title = {{Fast transient microjets induced by hemispherical cavitation bubbles}},
volume = {767},
year = {2015}
}

@article{Karri2011,
author = {Karri, Badarinath and Pillai, Kiran S. and Klaseboer, Evert and Ohl, Siew Wan and Khoo, Boo Cheong},
doi = {10.1016/j.sna.2011.04.015},
issn = {09244247},
journal = {Sensors and Actuators, A: Physical},
number = {1},
pages = {151--163},
publisher = {Elsevier B.V.},
title = {{Collapsing bubble induced pumping in a viscous fluid}},
url = {http://dx.doi.org/10.1016/j.sna.2011.04.015},
volume = {169},
year = {2011}
}

@article{Lew2007,
author = {Lew, Kelly Siew Fong and Klaseboer, Evert and Khoo, Boo Cheong},
doi = {10.1016/j.sna.2006.03.023},
issn = {09244247},
journal = {Sensors and Actuators, A: Physical},
number = {1},
pages = {161--172},
title = {{A collapsing bubble-induced micropump: An experimental study}},
volume = {133},
year = {2007}
}

@article{Khoo2005,
author = {Khoo, B.C. and Klaseboer, E. and Hung, K.C.},
doi = {10.1016/j.sna.2004.08.008},
issn = {09244247},
journal = {Sensors and Actuators A: Physical},
number = {1},
pages = {152--161},
title = {{A collapsing bubble-induced micro-pump using the jetting effect}},
volume = {118},
year = {2005}
}

@article{Pain2012,
author = {Pain, Agn{\`{e}}s and {Hui Terence Goh}, Bing and Klaseboer, Evert and Ohl, Siew Wan and {Cheong Khoo}, Boo},
doi = {10.1063/1.3692749},
issn = {00218979},
journal = {Journal of Applied Physics},
number = {5},
title = {{Jets in quiescent bubbles caused by a nearby oscillating bubble}},
volume = {111},
year = {2012}
}

@article{Goh2014,
author = {Goh, B. H.T. and Ohl, S. W. and Klaseboer, E. and Khoo, B. C.},
doi = {10.1063/1.4870244},
issn = {10897666},
journal = {Physics of Fluids},
number = {4},
title = {{Jet orientation of a collapsing bubble near a solid wall with an attached air bubble}},
volume = {26},
year = {2014}
}

@article{Karri2012b,
author = {Karri, Badarinath and Ohl, Siew Wan and Klaseboer, Evert and Ohl, Claus Dieter and Khoo, Boo Cheong},
doi = {10.1103/PhysRevE.86.036309},
issn = {15393755},
journal = {Physical Review E - Statistical, Nonlinear, and Soft Matter Physics},
number = {3},
pages = {1--11},
title = {{Jets and sprays arising from a spark-induced oscillating bubble near a plate with a hole}},
volume = {86},
year = {2012}
}

@article{Coussios2008,
author = {Coussios, Constantin C. and Roy, Ronald A.},
doi = {10.1146/annurev.fluid.40.111406.102116},
isbn = {9780824307400},
issn = {00664189},
journal = {Annual Review of Fluid Mechanics},
pages = {395--420},
title = {{Applications of acoustics and cavitation to noninvasive therapy and drug delivery}},
volume = {40},
year = {2008}
}

@article{Yusof2016,
author = {Yusof, Nor Saadah Mohd and Babgi, Bandar and Alghamdi, Yousef and Aksu, Mecit and Madhavan, Jagannathan and Ashokkumar, Muthupandian},
doi = {10.1016/j.ultsonch.2015.06.013},
issn = {18732828},
journal = {Ultrasonics Sonochemistry},
pages = {568--576},
pmid = {26142078},
publisher = {Elsevier B.V.},
title = {{Physical and chemical effects of acoustic cavitation in selected ultrasonic cleaning applications}},
url = {http://dx.doi.org/10.1016/j.ultsonch.2015.06.013},
volume = {29},
year = {2016}
}

@article{goh2013low,
  title={A low-voltage spark-discharge method for generation of consistent oscillating bubbles},
  author={Goh, Bing Hui Terence and Oh, YDA and Klaseboer, Evert and Ohl, Siew-Wan and Khoo, Boo Cheong},
  journal={Review of Scientific Instruments},
  volume={84},
  number={1},
  year={2013},
  publisher={AIP Publishing}
}

@article{mancuso2020critical,
  title={A critical review of the current technologies in wastewater treatment plants by using hydrodynamic cavitation process: principles and applications},
  author={Mancuso, Giuseppe and Langone, Michela and Andreottola, Gianni},
  journal={Journal of Environmental Health Science and Engineering},
  volume={18},
  number={1},
  pages={311--333},
  year={2020},
  publisher={Springer}
}

@article{lindau2003cinematographic,
  title={Cinematographic observation of the collapse and rebound of a laser-produced cavitation bubble near a wall},
  author={Lindau, Olgert and Lauterborn, Werner},
  journal={Journal of Fluid Mechanics},
  volume={479},
  pages={327--348},
  year={2003},
  publisher={Cambridge University Press}
}

@article{chahine1977interaction,
  title={Interaction between an oscillating bubble and a free surface},
  author={Chahine, GL},
  journal={Journal of Fluids Engineering},
  year={1977}
}

@inproceedings{gisbon1980growth,
  title={Growth and collapse of cavitation bubbles near flexible boundaries},
  author={Gisbon, DC and Blake, JR},
  booktitle={Australasian Conference on Hydraulics and Fluid Mechanics (7th: 1980: Brisbane, Qld.)},
  pages={283--286},
  year={1980},
  organization={Institution of Engineers, Australia Barton, ACT}
}

@article{maxworthy1972structure,
  title={The structure and stability of vortex rings},
  author={Maxworthy, T},
  journal={Journal of Fluid Mechanics},
  volume={51},
  number={1},
  pages={15--32},
  year={1972},
  publisher={Cambridge University Press}
}

@article{mnich2024single,
  title={Single cavitation bubble dynamics in a stagnation flow},
  author={Mnich, Dominik and Reuter, Fabian and Denner, Fabian and Ohl, Claus-Dieter},
  journal={Journal of Fluid Mechanics},
  volume={979},
  pages={A18},
  year={2024},
  publisher={Cambridge University Press}
}

@article{heidary2024robust,
  title={Robust cavitation-based pumping into a capillary},
  author={Heidary, Z and Fan, Y and Mojra, A and Ohl, CD},
  journal={Physics of Fluids},
  volume={36},
  number={12},
  year={2024},
  publisher={AIP Publishing}
}

@article{poulain2015particle,
  title={Particle motion induced by bubble cavitation},
  author={Poulain, St{\'e}phane and Guenoun, Gabriel and Gart, Sean and Crowe, William and Jung, Sunghwan},
  journal={Physical review letters},
  volume={114},
  number={21},
  pages={214501},
  year={2015},
  publisher={APS}
}

@article{tomita1984collapse,
  title={Collapse of multiple gas bubbles by a shock wave and induced impulsive pressure},
  author={Tomita, Yukio and Shima, Akira and Ohno, Takashi},
  journal={Journal of applied physics},
  volume={56},
  number={1},
  pages={125--131},
  year={1984},
  publisher={American Institute of Physics}
}

@article{vogel1989optical,
  title={Optical and acoustic investigations of the dynamics of laser-produced cavitation bubbles near a solid boundary},
  author={Vogel, A and Lauterborn, Werner and Timm, R},
  journal={Journal of Fluid Mechanics},
  volume={206},
  pages={299--338},
  year={1989},
  publisher={Cambridge University Press}
}

@article{gonzalez2011cavitation,
  title={Cavitation bubble dynamics in a liquid gap of variable height},
  author={Gonzalez-Avila, Silvestre Roberto and Klaseboer, Evert and Khoo, Boo Cheong and Ohl, Claus-Dieter},
  journal={Journal of Fluid Mechanics},
  volume={682},
  pages={241--260},
  year={2011},
  publisher={Cambridge University Press}
}

@article{teran2018interaction,
  title={Interaction of particles with a cavitation bubble near a solid wall},
  author={Teran, Leonel A and Rodriguez, Sara A and La{\'\i}n, Santiago and Jung, Sunghwan},
  journal={Physics of Fluids},
  volume={30},
  number={12},
  year={2018},
  publisher={AIP Publishing}
}

@article{brujan2001dynamics,
  title={Dynamics of laser-induced cavitation bubbles near an elastic boundary},
  author={Brujan, Emil-Alexandru and Nahen, Kester and Schmidt, Peter and Vogel, Alfred},
  journal={Journal of Fluid Mechanics},
  volume={433},
  pages={251--281},
  year={2001},
  publisher={Cambridge University Press}
}

@article{arndt1981cavitation,
  title={Cavitation in fluid machinery and hydraulic structures},
  author={Arndt, Roger EA},
  journal={Annual Review of Fluid Mechanics},
  volume={13},
  number={1},
  pages={273--326},
  year={1981},
  publisher={Annual Reviews 4139 El Camino Way, PO Box 10139, Palo Alto, CA 94303-0139, USA}
}

@article{sarraf2022fundamentals,
  title={Fundamentals, biomedical applications and future potential of micro-scale cavitation-a review},
  author={Sarraf, Seyedali Seyedmirzaei and Talabazar, Farzad Rokhsar and Namli, Ilayda and Maleki, Mohammadamin and Aghdam, Araz Sheibani and Gharib, Ghazaleh and Grishenkov, Dmitry and Ghorbani, Morteza and Ko{\c{s}}ar, Ali},
  journal={Lab on a Chip},
  volume={22},
  number={12},
  pages={2237--2258},
  year={2022},
  publisher={Royal Society of Chemistry}
}

@article{philipp1998cavitation,
  title={Cavitation erosion by single laser-produced bubbles},
  author={Philipp, Andreas and Lauterborn, Werner},
  journal={Journal of fluid mechanics},
  volume={361},
  pages={75--116},
  year={1998},
  publisher={Cambridge University Press}
}

@article{plesset1949dynamics,
  title={The dynamics of cavitation bubbles},
  author={Plesset, Milton S},
  year={1949},
  journal={Journal of Applied Mechanics},
  publisher={American Society of Mechanical Engineers}
}

@inproceedings{poritsky1952collapse,
  title={The collapse or growth of a spherical bubble or cavity in a viscous fluid},
  author={Poritsky, H},
  booktitle={Proceedings of the First US National Congress on Applied Mechanics},
  pages={813--821},
  year={1952},
  organization={New York}
}

\end{document}